\documentclass{article}
\usepackage[utf8]{inputenc}
\title{Numerical investigation of minimum drag profiles in laminar flow using deep learning surrogates}
\author{Li-Wei Chen, Berkay Alp Cakal, Xiangyu Hu, Nils Thuerey }
\date{September 2020}

\usepackage{amsmath}
\usepackage[numbers,sort&compress]{natbib}
\usepackage{graphicx}
\usepackage{epstopdf, epsfig}
\usepackage{xcolor}
\usepackage{bm}
\usepackage{subcaption}
\usepackage{float}
\usepackage{multirow}
\usepackage{booktabs}

\usepackage{color}
\definecolor{nCol}{rgb}{0.6, 0.2, 0.1}
\definecolor{lCol}{rgb}{0.2, 0.7, 0}
\definecolor{zCol}{rgb}{0.5, 0.6, 1.0}
\definecolor{todoCol}{rgb}{0.8, 0.1, 0}

\begin{document}

\maketitle

\begin{abstract}
Efficiently predicting the flowfield and load in aerodynamic shape optimisation remains a highly challenging and relevant task. Deep learning methods have been of particular interest for such problems, due to their success for solving inverse problems in other fields. In the present study, U-net based deep neural network (DNN) models are trained with high-fidelity datasets to infer flowfields, and then employed as surrogate models to carry out the shape optimisation problem, i.e. to find a drag minimal profile with a fixed cross-section area subjected to a two-dimensional steady laminar flow. A level-set method as well as B{\'e}zier-curve method are used to parameterise the shape, while trained neural networks in conjunction with automatic differentiation are utilized to calculate the gradient flow in the optimisation framework. The optimised shapes and drag force values calculated from the flowfields predicted by DNN models agree well with reference data obtained via a Navier-Stokes solver and from the literature, which demonstrates that the DNN models are capable of predicting not only flowfield but also yield satisfactory aerodynamic forces. This is particularly promising as the DNNs were not specifically trained to infer aerodynamic forces. In conjunction with the fast runtime, the DNN-based optimisation framework shows promise for general aerodynamic design problems.
\end{abstract}

\section{Introduction}
\noindent Owing to the importance in a wide range of fundamental studies and
industrial applications, a significant effort has been made 
to study the shape optimisation for minimising aerodynamic drag over a bluff body \citep{Bushnell1991annurev, Bushnell2003aircraft}. The deployment of computational fluid dynamics tools has played an important role in these optimisation problems \citep{Thevenin2008book}. 
While a direct optimisation via high-fidelity computational fluid dynamics (CFD) models gives reliable results, 
the high computational cost of each simulation, e.g., for Reynolds-averaged Navier-Stokes formulations, 
and the large amount of evaluations needed, lead to assessments that such optimisations are still not feasible for the practical engineering \citep{SKINNER2018933}.
When considering gradient-based optimisation, the adjoint method provides an effective way to calculate the gradients of an objective function w.r.t. design variables and alleviates the computational workload greatly \citep{Jameson1988control, Giles2000, economon_viscous_2013, kline_multi-objective_2016, zhou_efficient_2016}, but the number of required adjoint CFD 
simulations is typically still
prohibitively expensive when multiple optimisation objectives are considered \citep{Mueller2019}. 
In gradient-free methods (e.g. genetic algorithm), the computational cost rises dramatically as the number of design variables is increased, especially when the convergence requirement is tightened \citep{doi:10.3166/remn.17.103-126}. 
Therefore, 
advances in terms of surrogate-based optimisation are of central importance for both gradient-based and gradient-free optimisation methods \citep{QUEIPO20051, Sun_Wang2019reivew_surrogate}.

Recently, state-of-the-art deep learning methods and architectures have been successfully developed to achieve fast prediction of fluid physics. 
Among others, 
\citet{Bhatnagar_2019} developed convolutional neural network (CNN) method for aerodynamics flowfields, while others studied the predictability of laminar flows \citep{chen2019unet}, employed 
graph neural networks to predict transonic flows \citep{icml2020_6414},
or learned reductions of numerical errors in PDE fluid solvers \citep{um2020sol}.
For the inference of Reynolds-averaged Navier–Stokes (RANS) solutions, 
a U-net based deep learning model was proposed and shown to be significantly faster than a conventional CFD solver \citep{thuerey2018deep}.
These promising achievements open up new possibilities of applying DNN-based flow solvers in the aerodynamic shape optimisation.
In the present study we focus on evaluating the accuracy and performance of DNN-base surrogates in laminar flow regimes.

Modern deep learning methods are also giving new impetus to aerodynamic shape optimisation research. 
E.g., \citet{eismann2017shape} used a data-driven Bayesian approach to design optimization and generated object shapes with an improved drag coefficient. 
\citet{viquerat2019supervised}, evaluated quantitative predictions such as drag forces using a VGG-like convolutional neural network. To improve the surrogate-based optimisation, \citet{ShuChang2020} proposed a new sampling method for airfoils and wings based on a  generative adversarial network (GAN). \citet{renganathan2020enhanced} designed a surrogate-based framework 
by training a deep neural network (DNN)
that is used for gradient-based and gradient-free optimisations. In these studies, the neural network is mainly trained to construct the mapping between shape parameters and the aerodynamic quantities (e.g. lift and drag coefficients), but no flowfield information can be obtained from the network models.
We instead demonstrate how deep learning models that were not specifically trained to infer the parameters to be minimized, can be used in optimisation problems.
The proposed deep learning model offers two advantages. First, the model is flexible as it predicts a full flowfield in a region of interest. Once trained, it can be used in different optimization tasks with multiple objectives. This is of particular importance when considering problems such as compressor/turbine blade wake mixing \citep{MichelassiChen2015dns}. Second, as the model is differentiable, it can be seamlessly integrated with deep learning algorithms \citep{de2018end,holl2019pdecontrol}.

To understand the mechanisms underlying drag reduction and to develop optimisation algorithms, analytical and computational work have been specifically performed for Stokes' flow and laminar steady flow over a body \citep{Pironneau1973, Pironneau1974, Glowinski1975, Glowinski1976, kimkim1995, katamine2005, Kondoh2012}.
As far back as 1970s, \citet{Pironneau1973} analysed the minimum drag shape for a given volume in Stokes flow, and later for the Navier-Stokes equations \citep{Pironneau1974}.
By using the adjoint variable approach, \citet{kimkim1995} investigated the minimal drag profile for a fixed cross-section area in the two-dimensional laminar flow with the Reynolds number range of $Re=1$ to $40$. More recently \citet{katamine2005} studied the same problem at two Reynolds numbers $Re=0.1$ and $Re=40$.
With theoretical and numerical approaches, \citet{Glowinski1975, Glowinski1976} looked for the axisymmetric profile of given area and smallest drag in a uniform incompressible laminar viscous flow at Reynolds numbers between 100 and $10^5$ and obtained a drag-minimal shape with a wedge of angle $90^{\circ}$ at the front end and a cusp rear end from an initial slender profile. 
Although the laminar flow regimes are well studied, due to the separation and nonlinear nature of the fluid, it can be challenging for surrogate models to predict the drag-minimal shape as well as aerodynamic forces. 
Moreover, with the {Reynolds} number approaching zero,
the flowfield experiences a dramatic change from a separated vortical flow towards a creeping flow, which poses additional difficulties to the learning task.
To our knowledge, no previous studies exist that 
target training a ``generalised model'' that performs well in such a Reynolds number range.
We investigate this topic and quantitatively assess the results in the context of deep learning surrogates.

In the present paper, we adopt an approach for the U-net based flowfield inference \citep{thuerey2018deep} and use the trained deep neural network as flow solver in the shape optimisation. In comparison to conventional surrogate models \citep{YONDO201823} and other optimisation work involving deep learning \citep{eismann2017shape, ShuChang2020, renganathan2020enhanced, viquerat2019supervised}, we make use of a generic model that infers flow solutions: in our case it produces fluid pressure and velocity as field quantities. I.e., given encoded boundary conditions and shape, the DNN surrogate produces
a flowfield solution, from which the aerodynamic forces are calculated. Thus, both the flowfield and aerodynamic forces can be obtained during the optimisation. 
As we can fully control and generate arbitrary amounts of high-quality flow samples, we can train our models in a fully supervised manner. 
We use the trained DNN models in the shape optimisation to find the drag minimal profile in the two-dimensional steady laminar flow regime for a fixed cross-section area, and evaluate results w.r.t. shapes obtained using a full Navier-Stokes flow solver in the same optimisation framework. 
We specifically focus on the challenging Reynolds number range from 1 to 40.
Both level-set and B{\'e}zier-curve based methods are employed for shape parameterisation. The implementation utilizes the automatic differentiation package of the \textit{PyTorch} package \citep{paszke2019pytorch}, so the gradient flow driving the evolution of shapes can be directly calculated \citep{Kraft2017gradientFlow}. 
Here DNN-based surrogate models show particular promise as they allow for a seamless integration into the optimisation algorithms that are commonly used for training DNNs.

The purpose of the present work is to demonstrate the capability of deep learning techniques for robust and efficient shape optimisation, and for achieving an improved understanding of the inference of the fundamental phenomena involved in low Reynolds number flows.
This paper is organized as follows. The mathematical formulation and numerical
method are briefly presented in \S \ref{sec:methodology}. The neural network architecture and training procedure will be described in \S \ref{sec:neural_network}. The detailed experiments and results are then given in \S \ref{sec:opt_results} and concluding remarks in \S \ref{sec:concluding_remarks}.

\section{Methodology}\label{sec:methodology}

\noindent We first explain and validate our approach for computing the fluid flow environment in which shapes should be optimised. Afterwards, we describe two different shape parameterisiations, a level-set and a B{\'e}zier curve based one, which we employ for our optimisation results.

\subsection{Numerical procedure}\label{subsec:numerical_procedure}
\begin{figure}
\centering
\includegraphics[width=.85\textwidth]
{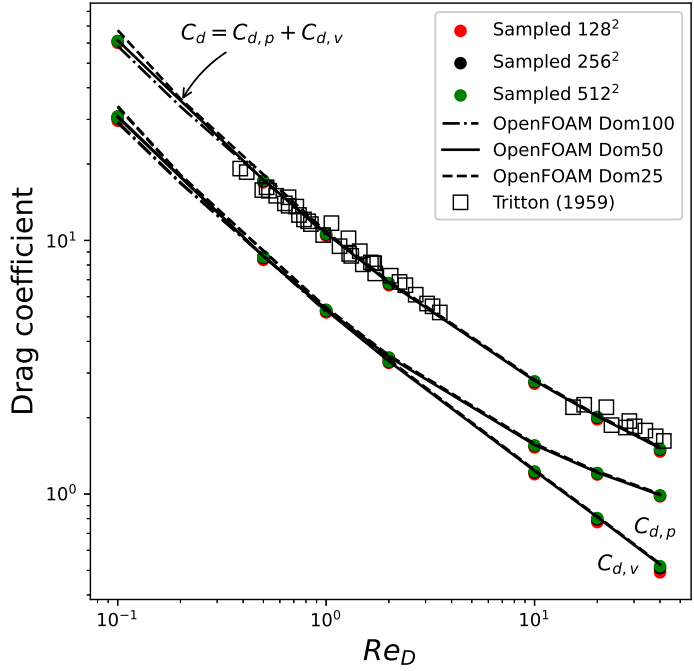}
\caption{Drag coefficients from $Re_D=0.1$ to 40. Surface integral values from \textit{OpenFOAM} simulations are plot in black curves. Results based on re-sampled points on Cartesian grids with the resolution of $128\times128$, $256\times256$ and $512\times512$ are plot with red, black and green circles, respectively. All data are compared with the experimental measurements by Tritton (1959), which are shown by blue squares.}
\label{fig:re_drag}
\end{figure}

\begin{figure}

\begin{subfigure}{.45\textwidth}
  \centering
  \includegraphics[width=\linewidth]{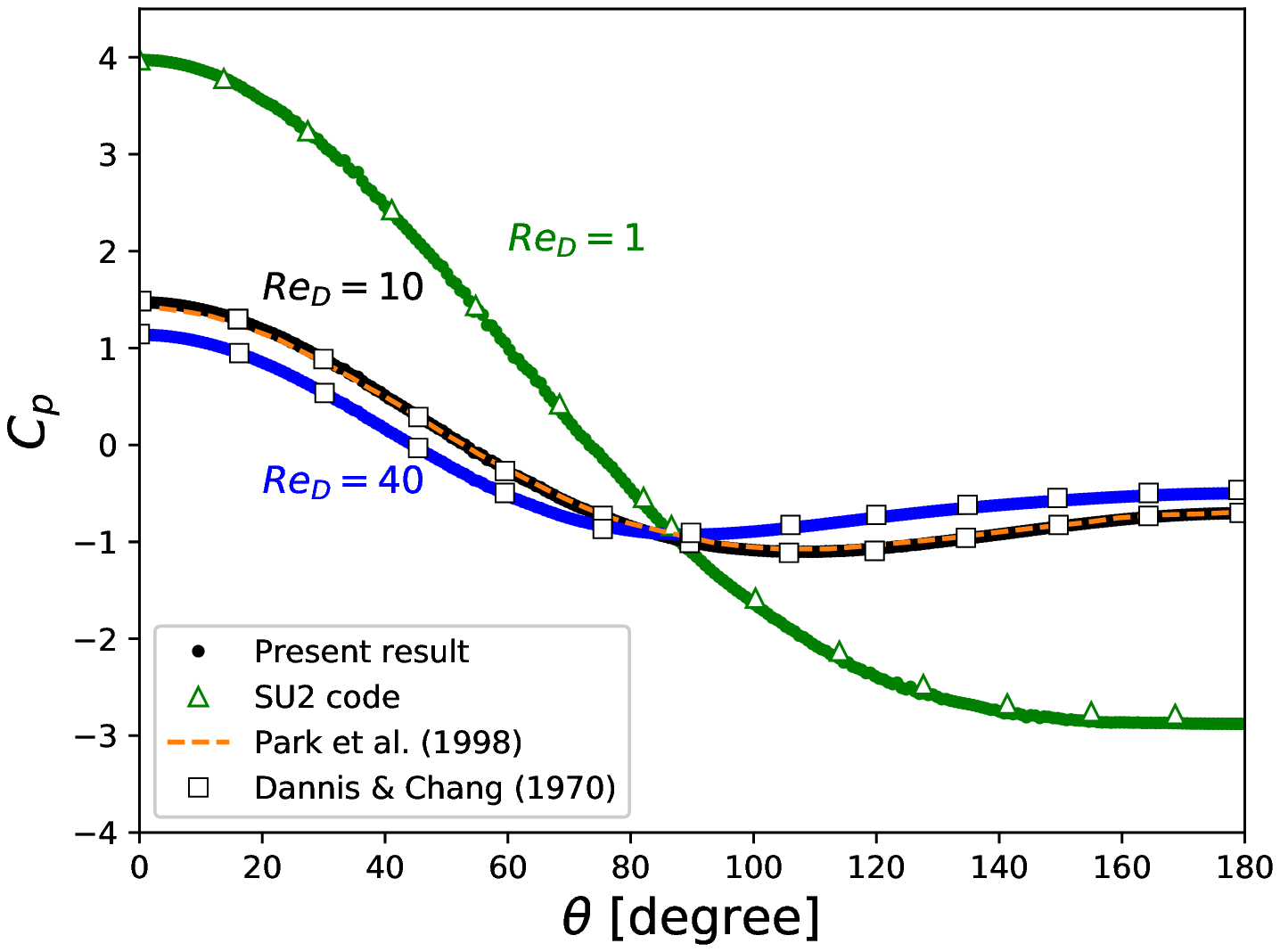}
  \caption{Pressure coefficient distribution}
  \label{fig:sub-first}
\end{subfigure}
\begin{subfigure}{.45\textwidth}
  \centering
  \includegraphics[width=\linewidth]{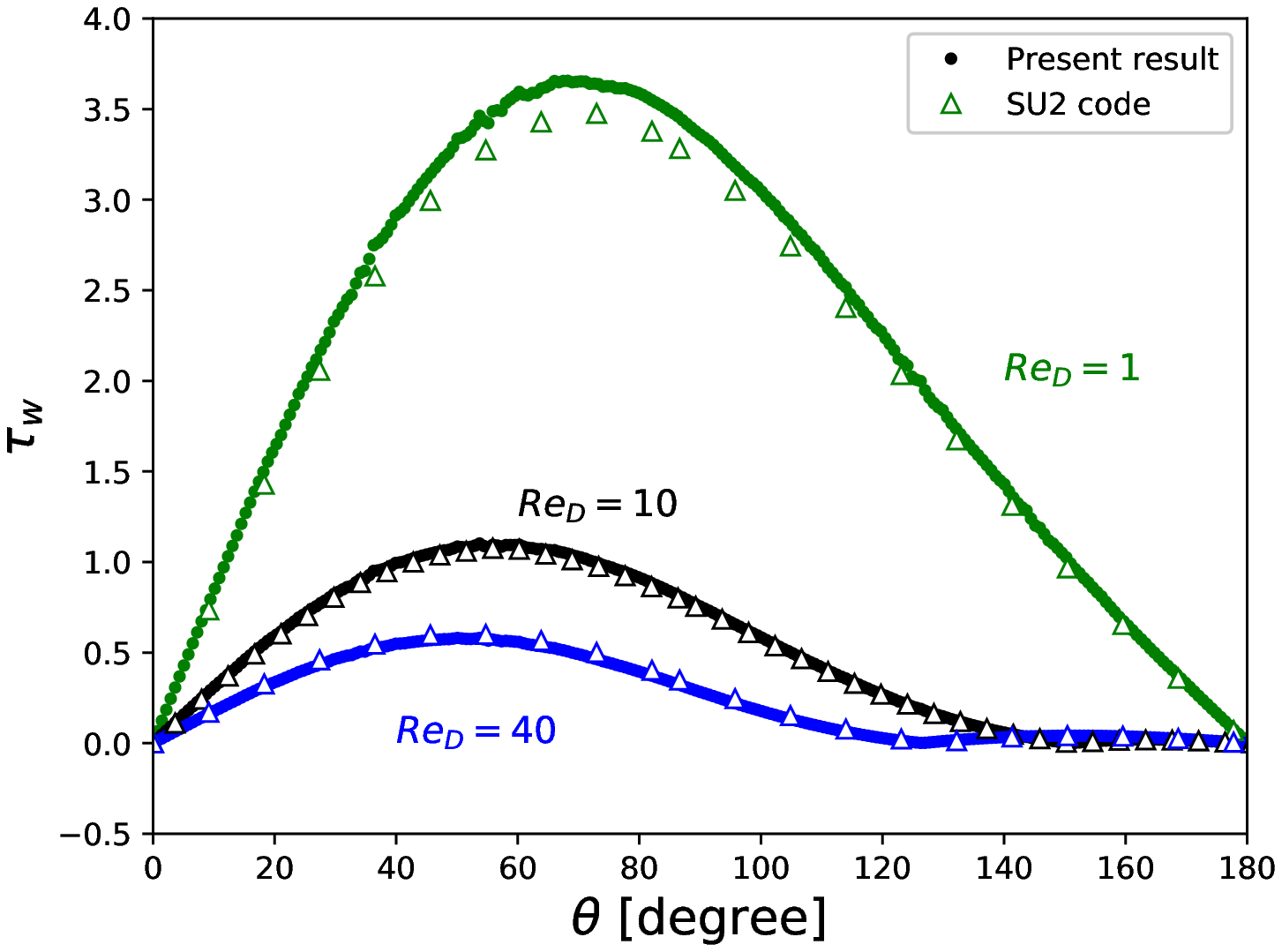}
  \caption{Wall shear stress distribution}
  \label{fig:sub-second}
\end{subfigure}

\caption{
\label{fig:cp}
	Pressure coefficient and wall shear stress distributions.}
\end{figure}

\noindent We consider two-dimensional incompressible steady laminar flows over profiles of given area and look for the minimal drag design. The profile is initialised with a circular cylinder and updated by utilizing steepest gradient descent as  optimisation algorithm.
The Reynolds number $Re_D$ in the present work is based on the diameter of the initial circular cylinder. It can be also interpreted that the the length scale is defined as the equivalent diameter for given area $S$ of an arbitrary shape, i.e. $D=2\sqrt{S/\pi}$. In the present work, $D\approx0.39424 [m]$ is used.

To calculate the flowfield around the profile at each iteration of the optimisation, two methods are employed in the present study. The first approach is a conventional steady solver of Navier-Stokes equations, i.e. \textit{simpleFoam} within the open-source package \textit{OpenFOAM} (from https://openfoam.org/). The second one is the deep learning model \citep{thuerey2018deep}, which is trained with flowfield datasets generated by \textit{simpleFoam}  that consists of several thousand
profiles at a chosen range of Reynolds numbers.
More details about the architecture of the neural network, data generation, training and performance will be discussed in \S \ref{sec:neural_network}.     

\textit{SimpleFoam} is a steady-state solver for incompressible, turbulent flow using the Semi-Implicit Method for Pressure Linked Equations (known as ``SIMPLE'') \citep{patankar1983}. The governing equations are numerically solved by a second-order finite volume method \citep{simplebook}. The unstructured mesh in the fluid domain is generated using open source code \textit{Gmsh} version 4.4.1. To properly resolve the viscous flow, the mesh resolution is refined near the wall of the profile and the minimum mesh size is set as $\sim6\times 10^{-3}D$, where $D$ is the equivalent circular diameter of the profile. The outer boundary, where the freestream boundary condition is imposed, is set as 50[m] ($\sim32D$) away from the wall (noted as 
``OpenFOAM DOM50''). The effects of domain size are assessed by performing additional simulations with domain size of 25[m] and 100[m] away from the wall (noted as ``OpenFOAM DOM25'' and ``OpenFOAM DOM100'', respectively). Here the drag coefficient $C_d$ is defined as the drag force divided by the projected length and dynamic head. As shown in figure \ref{fig:re_drag}, from $Re_D=0.1$ to $Re_D=40$, the total $C_d$ as well as the viscous $C_{d,v}$ and inviscid $C_{d,p}$ parts obtained from three different domains almost collapse. Although small differences when $Re_D<0.5$ are observed, the predictions in the interested range [1, 40] are consistent and not sensitive to the domain size. The computation runs for 6000 iterations to obtain a converged state.

To validate the setup, we compare our numerical results and literature data in terms of the surface pressure coefficient and wall shear. As sanity checks for the numerical setup, we also run 
\textit{SU2} 
\citep[see][]{su2code2016} with the same mesh for comparisons. 
Figure \ref{fig:cp}(a) shows the distribution of the surface pressure coefficient $(p_w-p_{\infty})/0.5\rho_{\infty}U_{\infty}^2$ at $Re_D=1$, 10 and 40. Here, $\theta$ is defined as the angle of the intersection of the horizontal line and the vector of the center to a local surface point, so that $\theta=0^{\circ}$ is the stagnation point in the upwind side and $\theta=180^{\circ}$ in the downwind side. Only half of the surface distribution is shown due to symmetry. The results agree well with the numerical results by \citet{DennisChang1970}, and the results for \textit{OpenFOAM} and \textit{SU2} collapse. In figure \ref{fig:cp}(b), the results for \textit{OpenFOAM} compare well with the one predicted by \textit{SU2}. 
The drag coefficients in the Reynolds numbers range from 0.1 to 40 agree well with the experimental data by Tritton et al. \citep{tritton_1959} in figure \ref{fig:re_drag}, which further supports that the current setup and the solver produce reliable data.

To facilitate neural networks with convolution layers, the velocity and pressure field from \textit{OpenFOAM} in the region of interest are re-sampled with a uniform Cartesian grid in a rectangular domain $[-1, 1]^2$ ($\approx[-1.27D, 1.27D]^2$). 
A typical resolution used in the present study is $128\times128$, corresponding to the grid size of $0.02D$. 
As also shown in figure \ref{fig:re_drag}, the effect of the resolution of re-sampling on the drag calculation has been studied. The details about the force calculation on Cartesian grids are given in \S \ref{subsec:level_set_rep}. Results with three different resolutions shown as colored symbols, i.e. $128^2$, $256^2$, and $512^2$, compare favourably with the surface integral values based on the original mesh in \textit{OpenFOAM}. Therefore, re-sampled fields on the $128\times128$ grid will be used in the deep learning framework and optimisation unless otherwise noted.

\subsection{Shape parameterisation} 
\subsubsection{Level-set method}\label{subsec:level_set_rep}
\noindent The level set method proposed by \citet{Osher1988} is a technique that tracks an interface implicitly and has been widely used in fluid physics, image segmentation, computer vision as well as shape optimisation \cite[]{Sethian1999book, Sethian2003annurev, Baeza2008}.
The level set function $\phi$ is a higher dimensional auxiliary scalar function, the zero-level set contour of which is the implicit representation of a time-dependent surface $\Gamma(t)=\{\boldsymbol{x}:\phi(\boldsymbol{x})=0\}$. Here, let $\mathcal{D}\in \mathcal{R}^\mathcal{N}$ be a reference domain, $\boldsymbol{x}\in \mathcal{D}$ and $\Omega$ is a body created by the enclosed surface $\Gamma$. Specifically in the present study, the domain $\mathcal{D}$ is referred to the sampled Cartesian grid in the rectangular region, and $\mathcal{N}$ is 2 as we focus on two-dimensional problems. The level set function $\phi$ is defined by a signed distance function (SDF):
\begin{equation}
    \phi=\begin{cases} 
      -d(\Gamma(t)) & \boldsymbol{x}\in \Omega \\
      0 & \boldsymbol{x}\in\partial \Omega\ (\text{on}\ \Gamma) \\
      d(\Gamma(t)) & \boldsymbol{x} \in \mathcal{D}-\Omega 
\end{cases}
\end{equation}
where $d(\Gamma(t))$ denotes the Euclidean distance from $\boldsymbol{x}$ to $\Gamma$. 

The arc length $c$ and area $S$ of the body are formulated as $c=\int_{\mathcal{D}} \delta_\epsilon (\phi) \mathopen|\nabla \phi \mathopen|ds $ and $S=\int_{\mathcal{D}}H_{\epsilon}(-\phi)ds$. To make the operators differentiable, in the above, we use smoothed Heaviside and Dirac Delta function $H_{\epsilon}(x)=\frac{1}{1+e^{-x/\epsilon}}$ and $\delta_{\epsilon}(x)=\partial_x\frac{1}{1+e^{-x/\epsilon}}$, respectively. $\epsilon$ is a small positive number and chosen as twice of the grid size \cite[]{Zahedi2010epsilon}.
Then, the aerodynamic forces due to pressure distribution and viscous effect are described as
\begin{equation}
\label{eqn:force_pressure}
  \boldsymbol{F}_{pressure}=\int_{\partial \Omega} ( p \boldsymbol{n}) dl = \int_{\mathcal{D}} (p\boldsymbol{n})\delta_\epsilon (\phi) \mathopen|\nabla \phi \mathopen|ds
\end{equation}
\begin{equation}
\label{eqn:force_viscous}
    \boldsymbol{F}_{viscous}=\int_{\partial \Omega} ( \mu \boldsymbol{n}\times\boldsymbol{\omega}) dl=\int_{\mathcal{D}} (\mu \boldsymbol{n}\times\boldsymbol{\omega})\delta_\epsilon (\phi) \mathopen|\nabla \phi \mathopen|ds.
\end{equation}
Here, $\boldsymbol{n}$ is the unit normal vector, $\boldsymbol{n}=\frac{\nabla \phi}{\mathopen| \nabla \phi \mathopen|}$, $p$ is the pressure, $\mu$ is the dynamic viscosity, and $\boldsymbol{\omega}=\nabla\times\boldsymbol{v}$ is the vorticity with $\boldsymbol{v}$ being the velocity. A nearest neighbour method is used to extrapolate values of pressure and vorticity inside the shape $\Omega$. Then, the drag force is considered as the loss in the optimisation, i.e. 
\begin{equation}
\label{eqn:loss_definition} 
    \mathcal{L}=\boldsymbol{F}_{pressure}\cdot\hat{\boldsymbol{i}}_x+\boldsymbol{F}_{viscous}\cdot\hat{\boldsymbol{i}}_x
\end{equation}
where $\hat{\boldsymbol{i}}_x$ is the unit vector in the direction of $x$ axis.

The minimisation of equation (\ref{eqn:loss_definition}) is solved by the following equation:
\begin{equation}
\label{eqn:levelset_transport}
    \frac{\partial \phi} {\partial \tau} + V_n \mathopen| \nabla \phi \mathopen| =0
\end{equation}
Here, the normal velocity is defined as $V_n=\frac{\partial \mathcal{L}}{\partial \phi}$. At every iteration, 
the Eikonal equation is solved numerically with the fast marching method to ensure
$\mathopen|\nabla \phi\mathopen|\approx1.0$ \cite[]{Sethian1999fastmarching}. Then, we have $\frac{\partial \phi} {\partial \tau} \propto  -\frac{\partial \mathcal{L}}{\partial \phi}$, which is a gradient flow that minimises the loss function $\mathcal{L}$ and drives the evolution of the profile \cite[]{He2007topological}. 
For a more rigorous mathematical analysis we refer to  \citet{Kraft2017gradientFlow}.
In the present work, the automatic differentiation fuctionality of \textit{PyTorch} is utilized to efficiently minimize equation (\ref{eqn:loss_definition}) via gradient descent.
Note that the level-set based surface representation and optimisation algorithm are relatively independent modules, and can be coupled with any flow solver, such as \textit{OpenFOAM} and \textit{SU2}, 
so long as the solver provides a re-sampled flowfield on the Cartesian grid (e.g. $128\times128$) at an iteration in the optimisation. We will leverage this flexibility by replacing the numerical solver with a surrogate model represented by a trained neural network below.

\subsubsection{B{\'e}zier-curve based parameterisation} 
\noindent B{\'e}zier curve based parametric shape parameterisation is a widely accepted technique in aerodynamic studies \citep{Gardner2003aiaa, Yang2018, Zhang2020}. This work utilizes two B{\'e}zier curves, representing upper and lower surfaces of the 
profile denoted with superscript \textit{k=\{u,l\}}. Control points $\mathbf{P}_{i}^{k} \in \mathcal{D}$ are the parameters of the optimisation framework. The B{\'e}zier curves are defined via following equation: 

\begin{equation}
\label{eqn:bezier_curve_def}
   \mathbf{B}^{k}(t)=\sum_{i=0}^{n} \binom{n}{i} t^{i} (1-t)^{n-i} \mathbf{P}_{i}^{k}
\end{equation}
where $t \in [0,1]$ denotes the sample points along the curves. First and the last control points of each curve share the same parameters to construct the closure $\Omega$ of the profile. 

A binary labeling of the Cartesian grid $\mathcal{D}$ is performed as
\begin{equation}
    \chi=\begin{cases} 
      1 & \boldsymbol{x}\in \Omega \\
      0 & \boldsymbol{x} \in \mathcal{D}-\Omega 
\end{cases}
\end{equation}
where $\chi$ is the binary mask of the profile and $\boldsymbol{x}$ is the coordinate of a point on the Cartesian grid.
The normal vector $\boldsymbol{n}$ is obtained via applying a convolution with a $3\times3$ Sobel operator kernel on $\chi$. Then, forces are calculated as
\begin{equation}
\label{eqn:force_pressure_bezier}
  \boldsymbol{F}_{pressure}= \sum_{i \in \mathcal{D}-\mathcal{\Omega}} (p\boldsymbol{n})_i \Delta l_i
\end{equation}
\begin{equation}
\label{eqn:force_viscous_bezier}
    \boldsymbol{F}_{viscous}=\sum_{i \in \mathcal{D}-\mathcal{\Omega}} (\mu\boldsymbol{n}\times\boldsymbol{\omega})_i \Delta l_i
\end{equation}
where $i$ is the index of a point outside the profile and $\Delta l_i$ is the grid size at the point $i$. Thereby, drag $\mathcal{L}$ is calculated using equation (\ref{eqn:loss_definition}).
As for the level set representation,
the shape gradient $\frac{\partial \mathcal{L}}{\partial \mathbf{P}_{i}^{k}}$ is computed via automatic differentiation in order to drive the shape evolution to minimize $\mathcal{L}$.

\section{Neural network architecture and training procedure}\label{sec:neural_network}
\subsection{Architecture}
\noindent The neural network model is based on a U-Net architecture  \citep{10.1007/978-3-319-24574-4_28}, a convolutional network originally used for the fast and precise segmentation of images. 
Following the methodology of previous work \citep{thuerey2018deep},
we consider the inflow boundary conditions (i.e. $u_{\infty}$, $v_{\infty}$) and the shape of profiles (i.e. the binary mask) on the Cartesian grid $128\times 128$ as three input channels. In the encoding part, 7 convolutional blocks are used to transform the input (i.e. $128^2\times3$) into a single data point with 512 features. The decoder part of the network is designed symmetrically with another 7 layers in order to reconstruct the outputs with the desired dimension, i.e. $128^2\times3$, corresponding to the flowfield variables $[p, u, v]$ on the Cartesian grid $128\times128$. 
Leaky ReLU activation functions with a slope of 0.2 is used in the encoding layers, and regular ReLU activations in the decoding layers.

In order to assess the performance of the deep learning models, we have tested three different models with varying weight counts of 122k, 1.9m and 30.9m, respectively, which are later referred to as {\em small, medium} and {\em large}-scale networks.

\subsection{Dataset generation}\label{subsec:dataset_generation}

\begin{figure}
    \centering
    \begin{subfigure}{.5\textwidth}
        \centering
        \includegraphics[width=.999\textwidth]
        {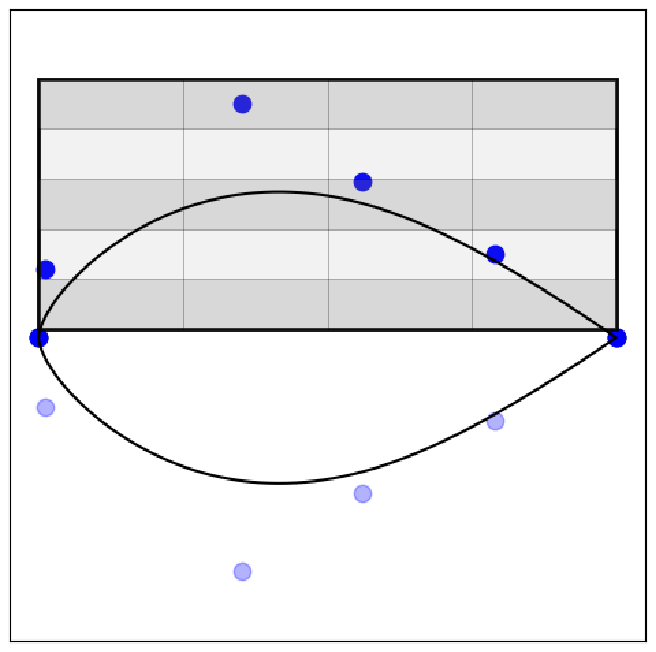}
        \caption{B{\'e}zier control points}
    \end{subfigure}%
    \begin{subfigure}{.5\textwidth}
        \centering
        \includegraphics[width=.999\textwidth]
        {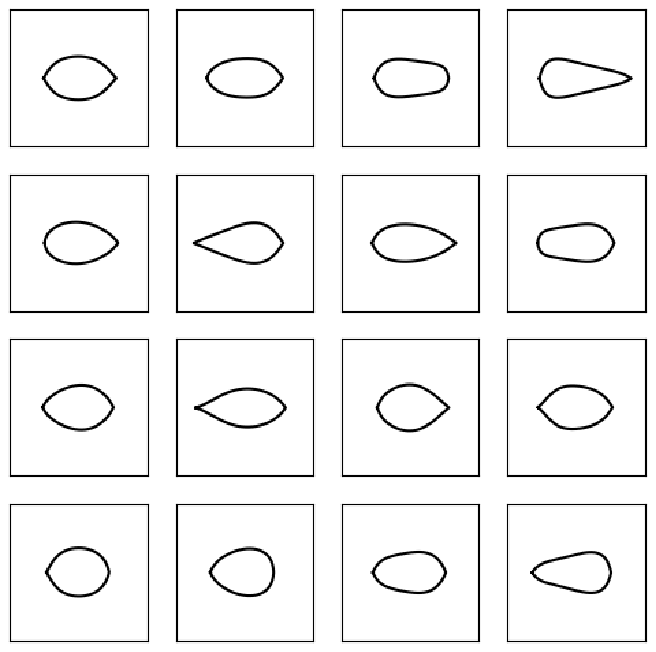}
        \caption{Randomly generated shapes}
    \end{subfigure}
    
    \caption{Shape generation using two B{\'e}zier curves. The region of interest is divided into 4 columns, and each column-wise region is further split into 5 sub-regions.}
    \label{fig:bezier_shape_generation}
\end{figure}

\noindent For the training dataset,
it is important to have a comprehensive coverage of the space of targeted solutions. 
In the present study, we utilize the parametric B{\'e}zier curve defined by equation (\ref{eqn:bezier_curve_def}) to generate randomized symmetric shape profiles subject to a fixed area constraint $S$.

To parameterise the upper surface of the profile, two points at the leading and trailing edges are fixed and 4 control points are positioned in different regions. As depicted in figure \ref{fig:bezier_shape_generation}(a), the region of interest is divided into 4 columns separated by the border lines, and each control point of the upper B{\'e}zier curve is only allowed to be located within its corresponding column-wise region. Each column-wise region is further split into 5 sub-regions to produce diversified profiles. 
The subregions give $5^{4}=625$ possible permutations, with control points being placed randomly in each subregion. 
This procedure is repeated for 4 times, in total it produces $4*625=2500$ B{\'e}zier curves. 
Figure \ref{fig:bezier_shape_generation}(b) shows some examples from this set.

Based on these 2500 geometries, we then generate three sets of training data, as summarised in table \ref{tab:dataset_details}. 

(1) We run \textit{OpenFOAM} with fixed $Re_D=1$ for all of the 2500 profiles to  obtain 2500 flowfields, denoted as ``Dataset-1''.

(2) The second dataset is similar but all of the 2500 simulations are conducted at $Re_D=40$ (``Dataset-40''). 

(3) The third dataset is generated to cover a continuous range of Reynolds numbers, in order to capture a space of solutions that not only varies over the immersed shapes, but additionally captures a dimensions of varying flow physics with respect to a chosen Reynolds number.
For this, we run a simulation by randomly choosing a profile $\Omega^*_i$ among 2500 geometries and a Reynolds number in the range of $Re^*_D \in [0.5, 42.5]$. 
As we know that drag scales logarithmically w.r.t. Reynolds number, we similarly employ a logarithmic sampling for the Reynolds number dimension.
We use a uniform distribution random variable $\kappa \in [\log{0.5}, \log{42.5}]$, leading to a $Re^*_D=10^{\kappa}$ uniformly distributed in log scale. 
In total we have obtained 8640 flowfields, which we refer to as ``Dataset-Range''. 
With this size of the training dataset, the model performance converges to a stable prediction accuracy for training and validation losses,
as shown in appendix \ref{appA}.

As shown in figure \ref{fig:randomMapMethod-III}, the distribution of all the flowfield samples from ``Dateset-Range'' on the $\Omega^*_i-Re^*_D$ map, with $Re^*_D$ in log scale. It worth noting that there are 2053 flowfield samples in the range of $Re^*_D \in [0.5, 1.5]$ which are shown in red, 819 samples with $Re^*_D \in [8,12]$ colored in green and 195 samples with $Re^*_D \in [38, 42]$ in blue.

\begin{table}
  \begin{center}
\def~{\hphantom{0}}
  \begin{tabular}{lccc}
      \toprule
      Name    & \# of flowfields & Re & NN models\\
      \midrule 
      Dataset-1    & 2500  & 1 & small, medium \& large\\
      Dataset-40      & 2500 & 40	 & small, medium \& large\\
      Dataset-Range     & 8640 & 0.5-42.5 & large\\
      \bottomrule
  \end{tabular}
  \caption{Three datasets for training the neural network models.}
  \label{tab:dataset_details}
  \end{center}
\end{table}

\begin{figure}
\centering
\includegraphics[width=1\textwidth]
{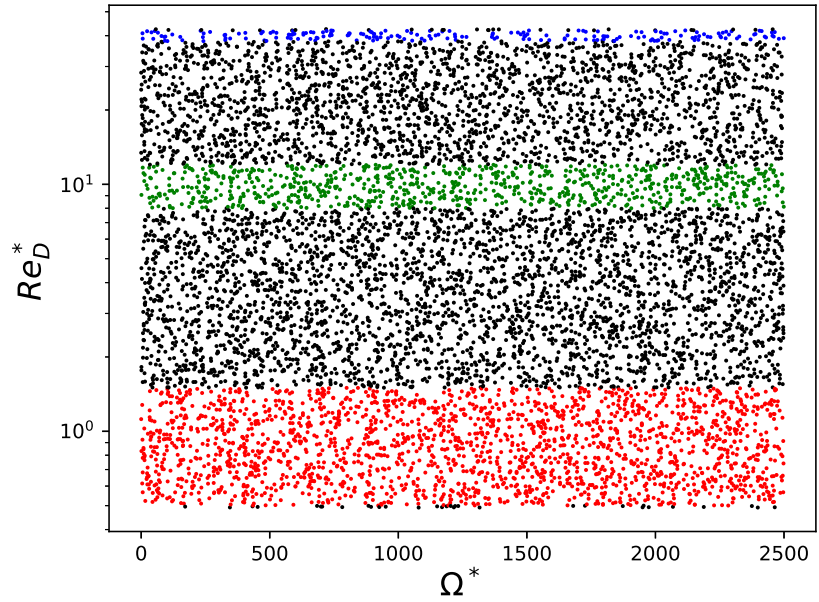}
\caption{Distribution of flowfield samples from ``Dataset-Range'' on the $\Omega^*_i-Re^*_D$ map. The indices of geometries $\Omega^*_i$ are from 0 to 2499. The red symbols denote the flowfield samples with $Re^*_D \in [0.5, 1.5]$, the green ones with $Re^*_D \in [8,12]$ and the blue ones with $Re^*_D \in [38, 42]$.}
\label{fig:randomMapMethod-III}
\end{figure}

\subsection{Pre-processing}

\noindent Proper pre-processing of the data is crucial for obtaining a high inference accuracy from the trained neural networks. Firstly, the nondimensional flowfield variables are calculated by
\begin{equation*} 
\begin{aligned}
\hat{p}_i=(p_i-p_{i, mean})/U_{\infty, i}^2, \\
\hat{u}_i=u_i/U_{\infty, i}, \\
\hat{v}_i=v_i/U_{\infty, i}.
\end{aligned}
\end{equation*}
Here, $i$ denotes the $i$-th flowfield sample in the dataset, $p_{mean}$ the simple arithmetic mean pressure, and $U_{\infty}=\sqrt{u_{\infty}^2+v_{\infty}^2}$ the magnitude of the freestream velocity. 

As the second step, all input channels and target flowfield data in the training dataset are normalised to the range of $[-1, 1]$ in order to minimise the errors from limited precision in the training phase. To do so, we need to find the maximum absolute values for each flow variable in the entire training dataset, i.e. $|\hat{p}|_{max}$, $|\hat{u}|_{max}$ and $|\hat{v}_{max}|$. Similarly, the maximum absolute values of the freestream velocity components are $|u_{\infty}|_{max}$ and $|v_{\infty}|_{max}$. Then we get the final normalised flowfield variables in the following form:
\begin{equation*} 
\begin{aligned}
\Tilde{p}_i= \hat{p}_i/|\hat{p}|_{max}\\
\Tilde{u}_i= \hat{u}_i/|\hat{u}|_{max}\\
\Tilde{v}_i= \hat{v}_i/|\hat{v}|_{max}
\end{aligned}
\end{equation*}
and the normalised freestream velocities used for input channels are 
\begin{equation*} 
\begin{aligned}
\Tilde{u}_i= u_i/\mathbf{max}(|u_{\infty}|_{max}, 1\times10^{-18})\\
\Tilde{v}_i= v_i/\mathbf{max}(|v_{\infty}|_{max}, 1\times10^{-18})
\end{aligned}
\end{equation*}
The freestream velocities appear in the boundary conditions, on which the solution globally depends, and should be readily available spatially and throughout the different layers. Thus, freestream conditions and the shape of the profile are encoded in a $128^2\times3$ grid of values. The magnitude of the freestream velocity is chosen such that it leads to a desired Reynolds number.

\subsection{Training details}

\begin{figure}
\centering

\begin{subfigure}{.55\textwidth}
\centering
\includegraphics[width=\linewidth]{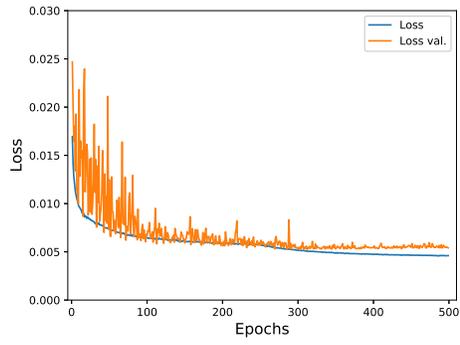}
\caption{Small-scale neural network}
\end{subfigure}

\begin{subfigure}{.55\textwidth}
\centering
\includegraphics[width=\linewidth]{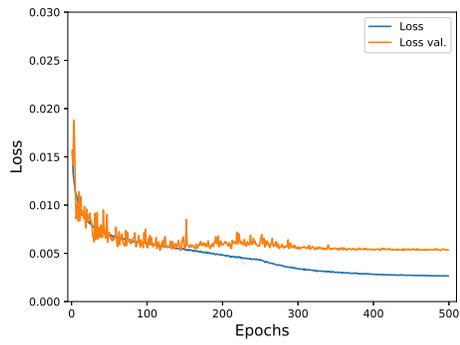}
\caption{Medium-scale neural network}
\end{subfigure}

\begin{subfigure}{.55\textwidth}
\centering
\includegraphics[width=\linewidth]{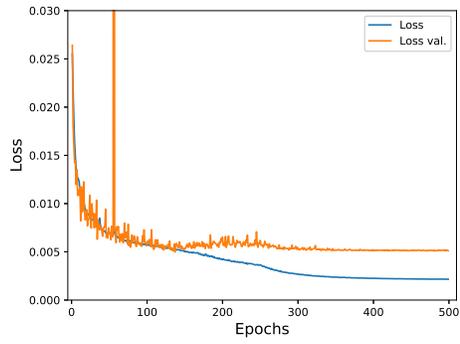}
\caption{Large-scale neural network}
\end{subfigure}

\caption{Training (in blue) and validation (in orange) losses of three different scales of models trained with ``Dataset-1''.} 
\label{fig:train_loss_re1}
\end{figure}

\begin{figure}
\centering

\begin{subfigure}{.55\textwidth}
\centering
\includegraphics[width=\linewidth]{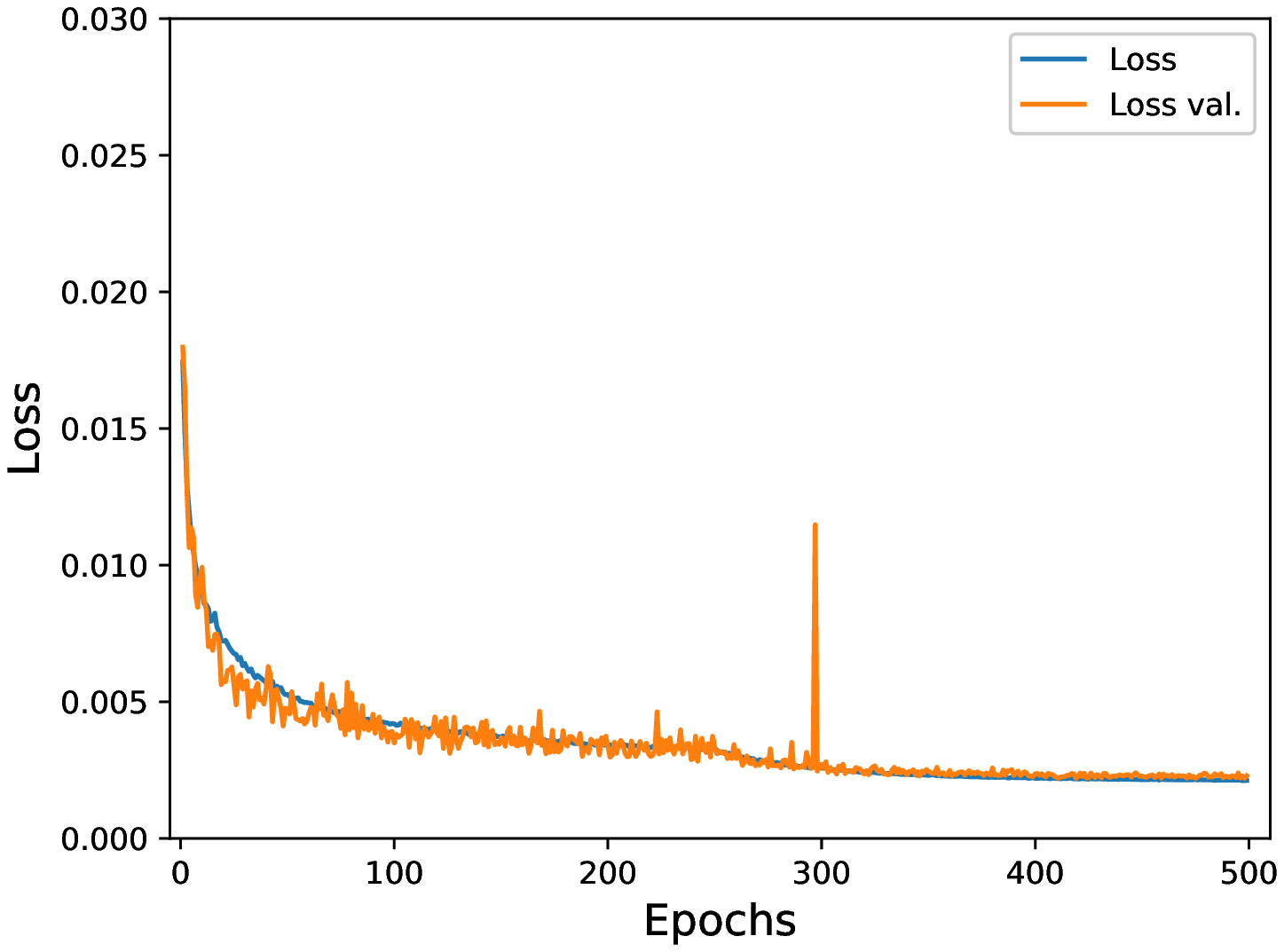}
\caption{Small-scale neural network}
\end{subfigure}

\begin{subfigure}{.55\textwidth}
\centering
\includegraphics[width=\linewidth]{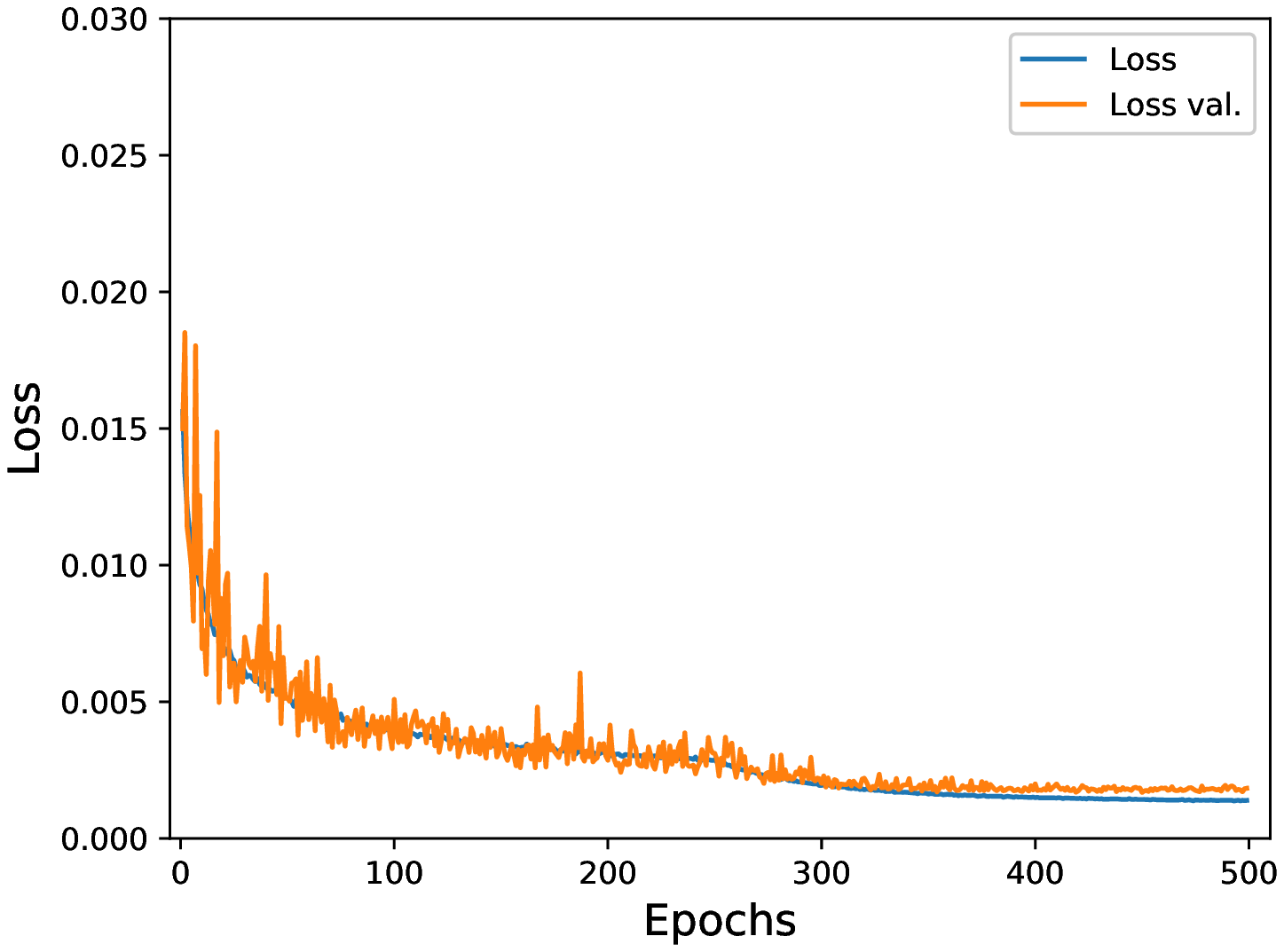}
\caption{Medium-scale neural network}
\end{subfigure}

\begin{subfigure}{.55\textwidth}
\centering
\includegraphics[width=\linewidth]{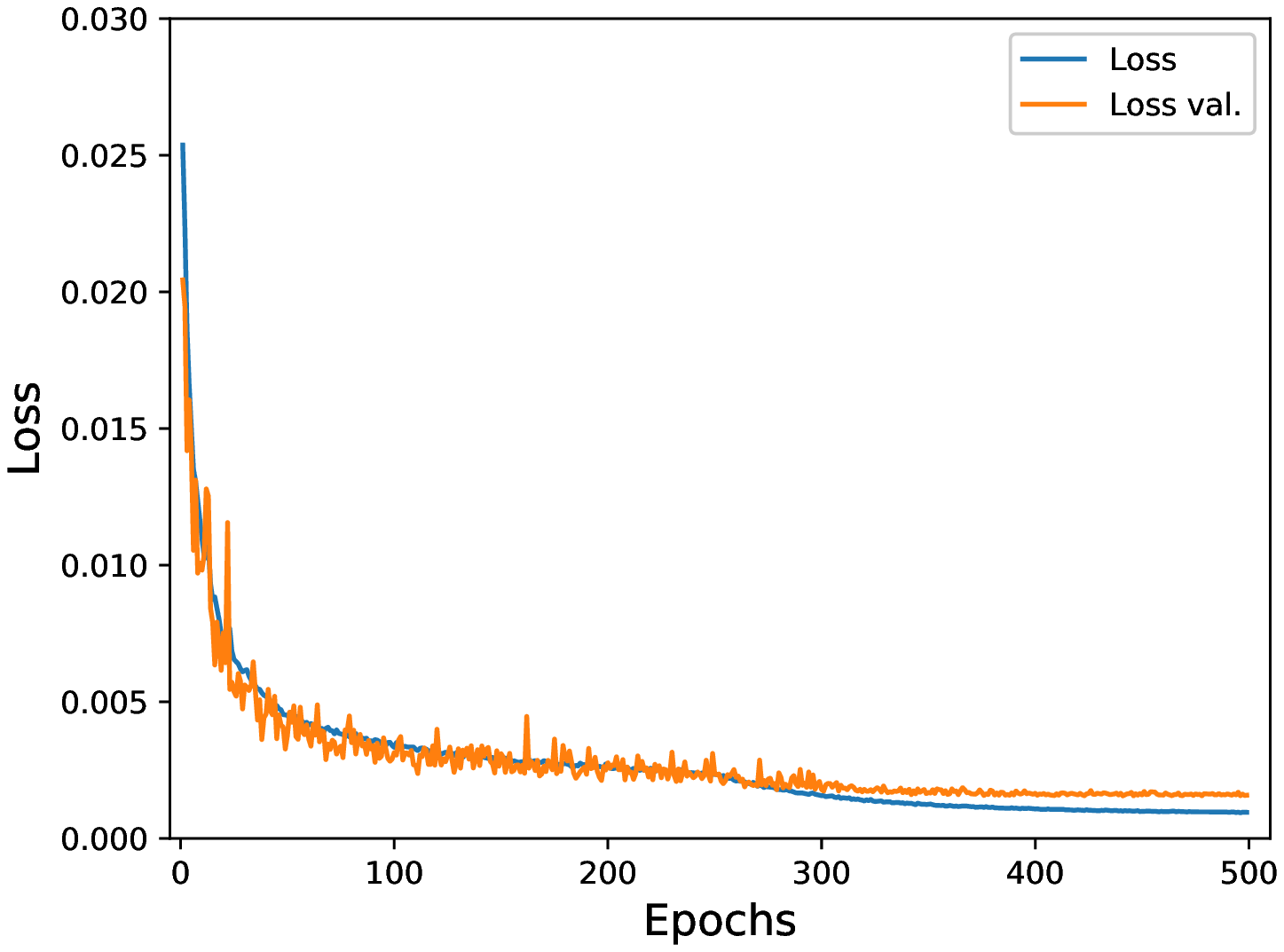}
\caption{Large-scale neural network}
\end{subfigure}

\caption{Training (in blue) and validation (in orange) losses of three different scales of models trained with ``Dataset-40''.}
\label{fig:train_loss_re40}
\end{figure}

\begin{figure}
\centering
\includegraphics[width=0.55\textwidth]
{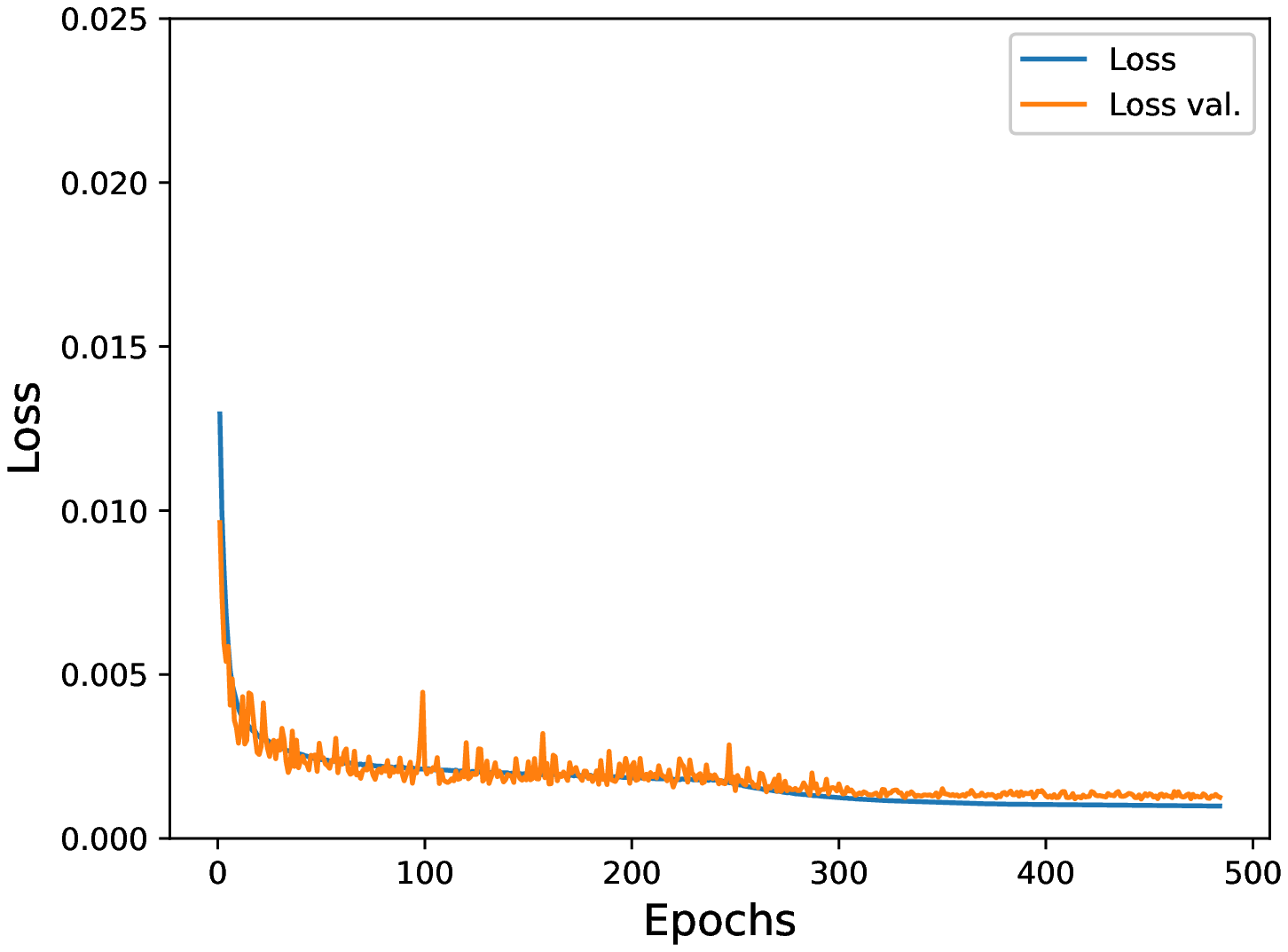}
\caption{Training (in blue) and validation (in orange) losses of large-scale model trained with ``Dataset-Range''.}
\label{fig:train_loss_ReGeneral}
\end{figure}

\noindent The neural network is trained with the Adam optimiser in \textit{PyTorch} \cite[]{kingma2014adam}. A $L_1$ difference $L_1=\mathopen|\mathbf{y}_{truth}-\mathbf{y}_{prediction}\mathopen|$ is used for the loss calculation. %
For most of the cases, the training runs converge after 100k iterations with a learning rate $6\times10^{-4}$ and a batch size of 10 (unless otherwise mentioned). An 80\% to 20\% split is used for training and validation sets, respectively. The validation set allows for an unbiased evaluation of the quality of the trained model during training, for example, to detect overfitting. In addition, as learning rate decay is used, the variance of the learning iterations gradually decreases, which lets the training process fine-tune the final state of the model.

Figure \ref{fig:train_loss_re1} shows the training and validation losses for three models that are trained using ``Dataset-1'',
i.e. small-scale, medium-scale and large-scale models, respectively. All the three models converge at stable levels of training and validation loss after 500 epochs. Looking at the training evolution for the small-scale model in figure \ref{fig:train_loss_re1}(a), numerical oscillation can be seen in the early stage of the validation loss history, which is most likely caused by the smaller number of free parameters in the small-scale network. In contrast, the medium- and large-scale models show a smoother loss evolution, and the gap between validation and training losses indicates  a slight overfitting as shown in figures \ref{fig:train_loss_re1}(b) and \ref{fig:train_loss_re1}(c). 
Although the training of the large-scale model exhibits a spike in the loss value at early stage
due to an instantaneous pathological configuration of mini-batch data and learned state, the network recovers, and eventually the converges to lower loss values. Similar spikes can be seen in some of the other training runs, and could potentially be removed via gradient-clipping algorithms, which we, however, did not find necessary to achieve reliable convergence.

Figure \ref{fig:train_loss_re40} presents the training and validation losses for three models trained with ``Dataset-40''.
Similarly, convergence can be achieved after 500 epochs. Compared to the training evolution at $Re_D=1$, the models $Re_D=40$ have smaller gaps between training and validation losses, indicating that the overfitting is less pronounced than for $Re_D=1$. We believe this is caused by the smoother and more diffusive flowfields at $Re_D=1$ (close to Stokes flow), in contrast to the additional complexity of the solutions at $Re_D=40$,
which already exhibit separation bubbles.

We use ``Dataset-Range'' to train the model for a continuous range of Reynolds numbers. 
As this task is particularly challenging, we directly focus on the large scale network that has 30.9m weights. 
To achieve better convergence for this case, we run 800k iterations with the batch size of 5, which leads to more than 485 epochs. As shown in figure \ref{fig:train_loss_ReGeneral}, training and validation losses converge to stable levels, and do not exhibit overfitting over the course of the training iterations. 
The final loss values are $1.01\times10^{-3}$ and $1.31\times10^{-3}$, respectively.

To summarise, having conducted the above-mentioned training, 
we obtain seven neural network models, i.e. models of three network sizes for ``Dataset-1'' and ``Dataset-40'' and a ranged model trained with ``Dataset-Range'' as list in table \ref{tab:dataset_details}. 
These neural networks will be used as surrogate models in the optimisation in the next section. 
We will also compare the results from neural network models with corresponding optimisations conducted with the \textit{OpenFOAM} solver, and evaluate the performance and accuracy of the optimisation runs.

\section{Shape optimisation results}\label{sec:opt_results}

\noindent The initial shape for the optimisation is a circular cylinder with a diameter $D\approx0.39424 [m]$. The integral value of the drag force using equation (\ref{eqn:loss_definition}) is adopted as the objective function. The mathematical formula of the optimisation for the shape $\Omega$ bounded by curve $\Gamma$, the surface of the profile, is expressed as

\begin{equation*}
\begin{aligned}
    \text{min}\ Drag({\Omega})\\
    \text{subject}\ \text{to}\ \text{Area}\ S(\Omega)=S_0 \\
    \text{Barycenter}\ \mathbf{b}(\Omega) = \frac{1}{S(\Omega)}\int_{\Omega} \boldsymbol{x}ds = (0,0)
\end{aligned}
\end{equation*}

For the level-set representation, the profile $\Omega$ is the region where $\phi\leq0$ and the constrained optimisation problem is solved as follows:

(1) Initialise level set function $\phi$ such that the initial shape (i.e. a circular cylinder) is corresponding to $\phi=0$.

(2) For a given $\phi$, calculate drag (i.e. loss $\mathcal{L}$) using Equations (\ref{eqn:force_pressure}-\ref{eqn:loss_definition}). Terminate if the optimisation converges, e.g. drag history reaches a statistically steady state.

(3) Calculate the gradient $\frac{\partial \mathcal{L}}{\partial \phi}$. Consider an unconstrained minimisation problem and solve equation (\ref{eqn:levelset_transport}) as follows:
\begin{equation*}
\phi^{n+1}\Longleftarrow\phi^{n} - \Delta \tau \frac{\partial \mathcal{L}}{\partial \phi} \mathopen| \nabla \phi \mathopen|
\end{equation*}

In practice, we update $\phi$ using the second-order Runge-Kutta method, and discretise the convection term with a first-order upwind scheme \cite[]{Sethian2003annurev}. 
We assume derivatives of the flowfield variables (i.e. pressure and velocity) are significantly smaller than 
those w.r.t. to the shape. Hence, we treat both fields as constants for each step of the shape evolution.
To ensure the correct search direction for optimisation, we use a relatively small pseudo time step $\Delta\tau$, which is calculated with a CFL number of 0.8.

(4) To ensure $\mathopen| \nabla \phi \mathopen| \approx 1$, a fast marching method is used to solve the Eikonal equation \cite[]{Sethian1999fastmarching}.

(5) The area of the shape $\Omega$ is obtained by $S=\int_{\mathcal{D}}H_{\epsilon}(-(\phi+\eta))ds$, where $\eta$ is an adjustable constant. We optimise $\eta$ such that $\mathopen| S-S_0\mathopen|<\epsilon$. Then, we update 
${\phi}^{n+1} \Longleftarrow {\phi}^{n+1} + \eta$.

(6) Check if the barycenter is at the origin: $\mathopen| \mathbf{b} - \mathbf{o}\mathopen|<\epsilon$. If not, solve equation (\ref{eqn:levelset_transport}) to update $\phi^{n+1}$ by replacing $V_n$ with a translating constant velocity so that the barycenter of the shape $\Omega$ moves towards the origin. Continue with (2).

While our main focus lies on level-set representations, the B{\'e}zier curve parameterisation with reduced degrees of freedom are used for comparison purposes. They highlight how differences in the shape parameterisation can influence the optimisation results. Thus, we include the Bezier parameterisation with very few control points and the level-set representation with a dense grid sampling as two extremes of the spectrum of shape representations.
When B{\'e}zier curves are used, the constrained optimisation defers from the above-mentioned loop in the following:
In (1-3), the coordinates of B{\'e}zier curve control points are used as the design variables to be initialised and updated.
In (5) and (6), the area of $\Omega$ and barycenter are calculated based on the region enclosed by the B{\'e}zier curve, where the binary mask of inner region is 1 and outer region is 0. %

In the optimisation experiments the flowfield solvers used are \textit{OpenFOAM} (as baseline), small, medium and large-scale neural network models, respectively. As additional validation for the optimisation procedure, we also compare to additional runs based on the B{\'e}zier curve parameterisation with a large-scale neural network model. If the flow solver is \textit{OpenFOAM}, \textit{Gmsh} is automatically called to generate an unstructured mesh based on the curve $\phi=0$ at every iteration. To update $\phi$ and calculate drag in Step (2), as aforementioned in \S\ref{subsec:numerical_procedure}, the flowfield variables are re-sampled on the $128\times128$ Cartesian grid. 

\subsection{Optimisation experiment at $Re_D=1$}

\begin{figure}

\begin{subfigure}{.45\textwidth}
\centering
\includegraphics[width=\linewidth]{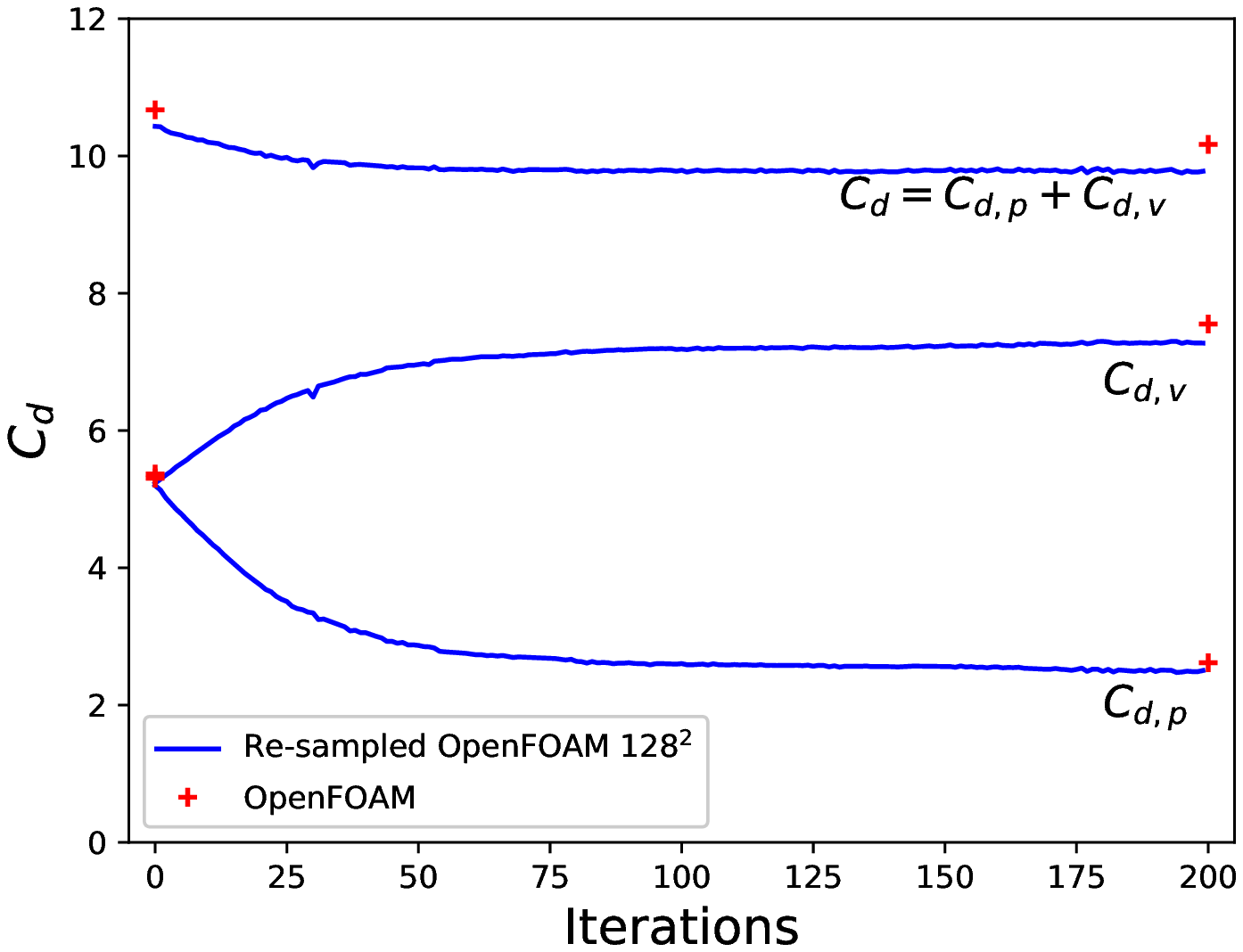}
\caption{OpenFOAM}
\end{subfigure}
\begin{subfigure}{.45\textwidth}
\centering
\includegraphics[width=\linewidth]{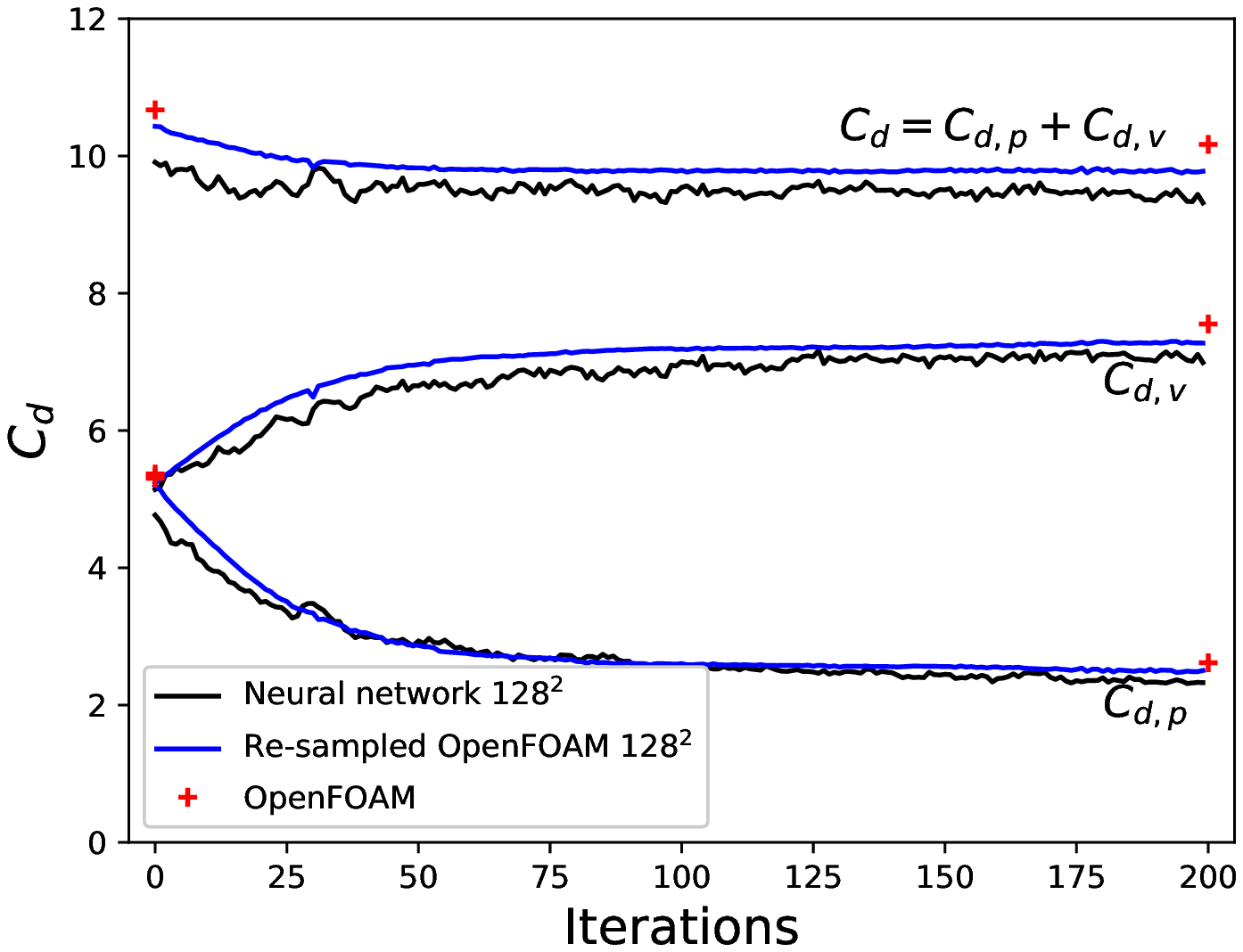}
\caption{Small-scale neural network}
\end{subfigure}

\begin{subfigure}{.45\textwidth}
\centering
\includegraphics[width=\linewidth]{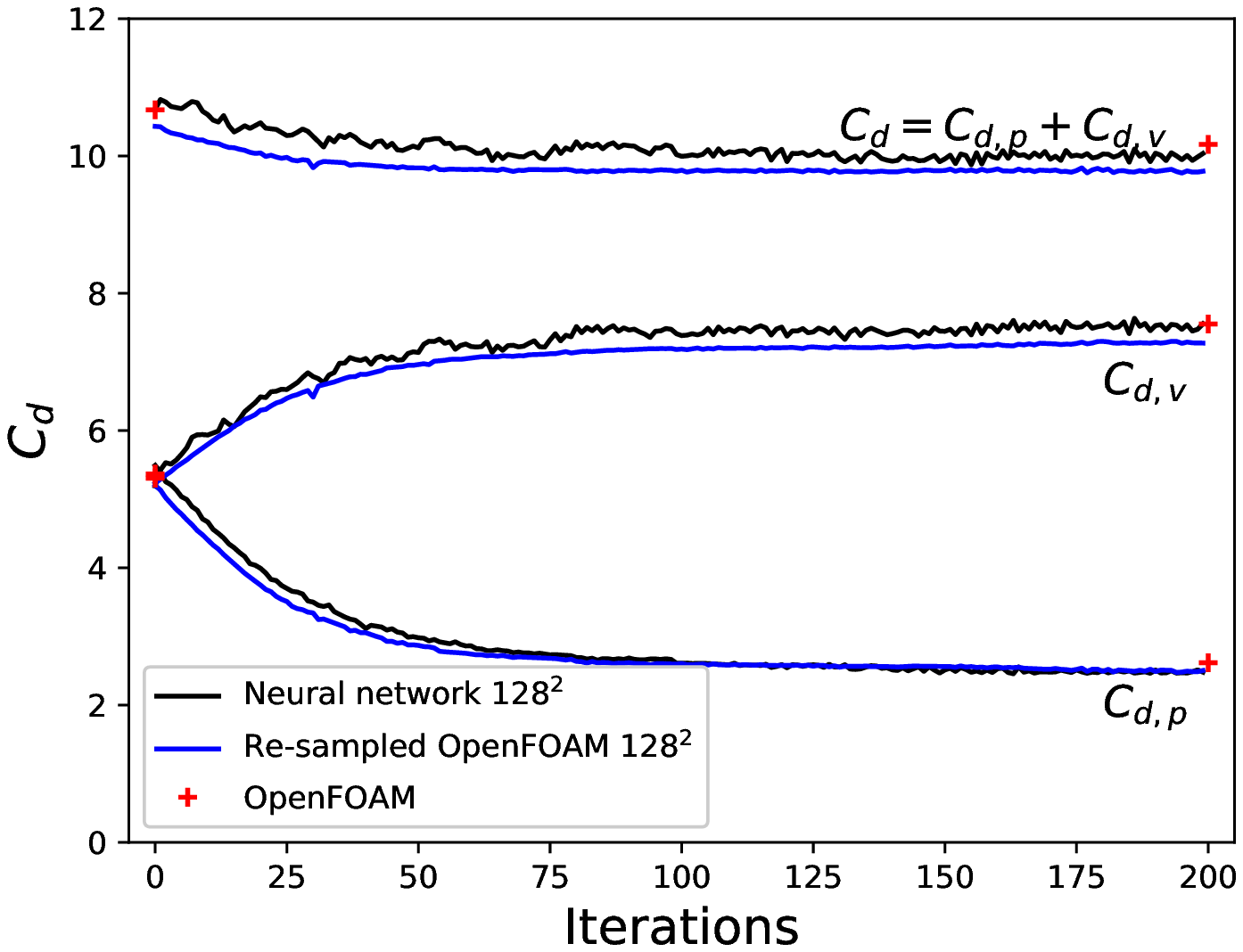}
\caption{Medium-scale neural network}
\end{subfigure}
\begin{subfigure}{.45\textwidth}
\centering
\includegraphics[width=\linewidth]{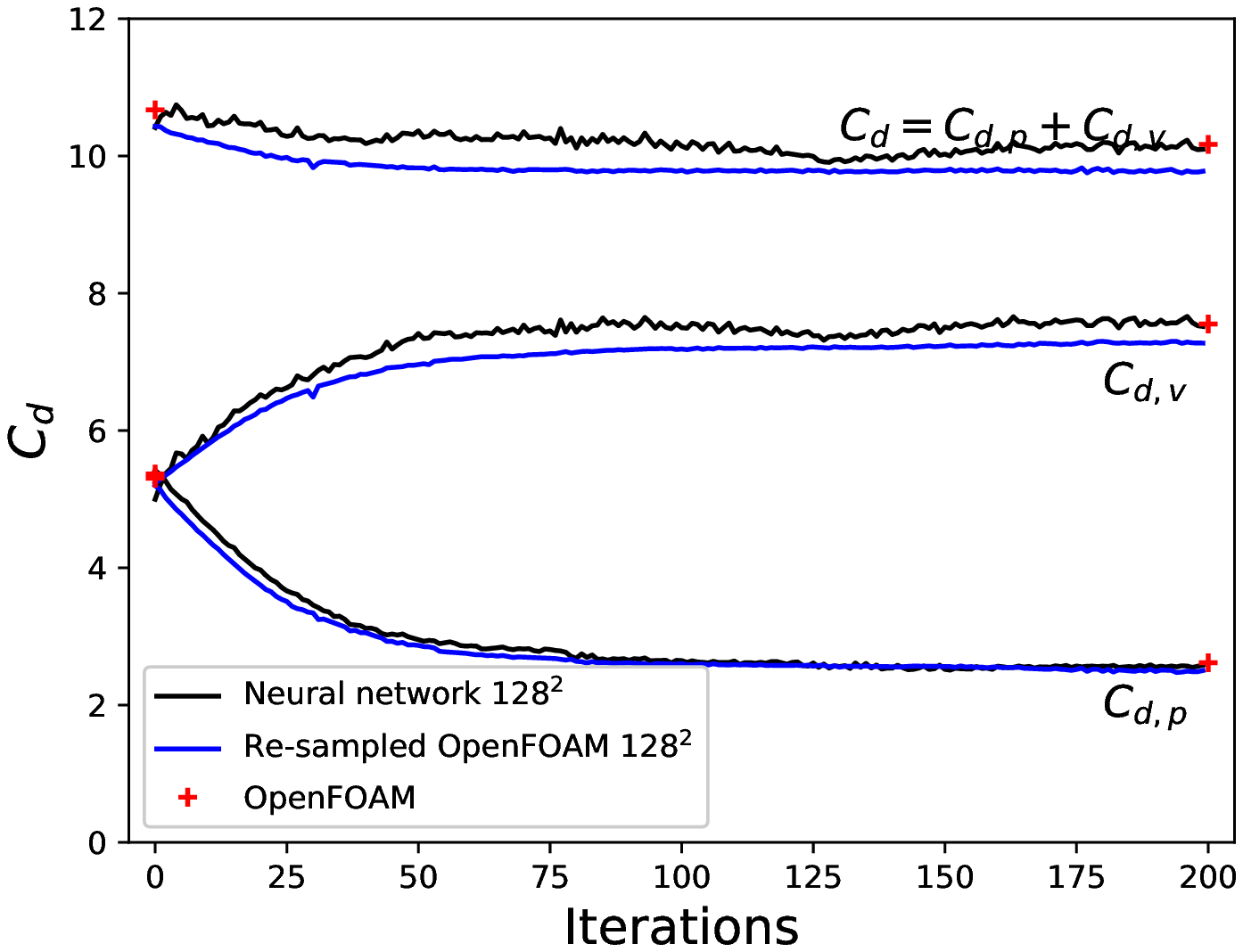}
\caption{Large-scale neural network}
\end{subfigure}

\caption{Optimisation histories at $Re_D=1$. The black solid lines denote the results using neural network models trained with ``Dataset-1'' and the blue solid lines denote the results from \textit{OpenFOAM}. Results calculated with the re-sampled flowfields on the $128\times128$ Cartesian grid are denoted by $128^2$. The red cross symbols represent the \textit{OpenFOAM}'s results obtained with its native postprocessing tool.}
\label{fig:drag_re1}
\end{figure}

\begin{figure}
\begin{subfigure}{.45\textwidth}
\centering
\includegraphics[width=\linewidth]{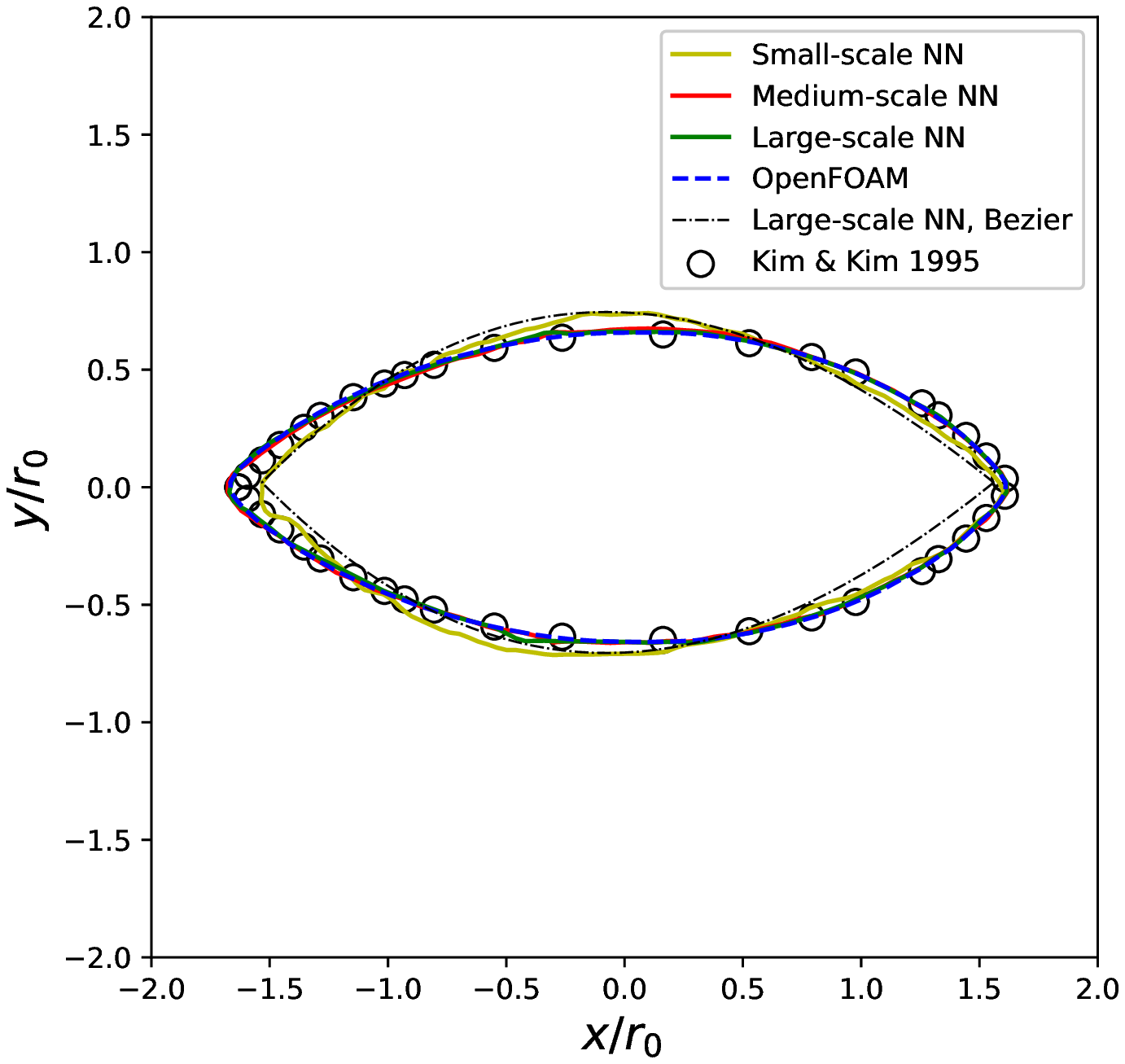}
\caption{}
\end{subfigure}
\begin{subfigure}{.45\textwidth}
\centering
\includegraphics[width=\linewidth]{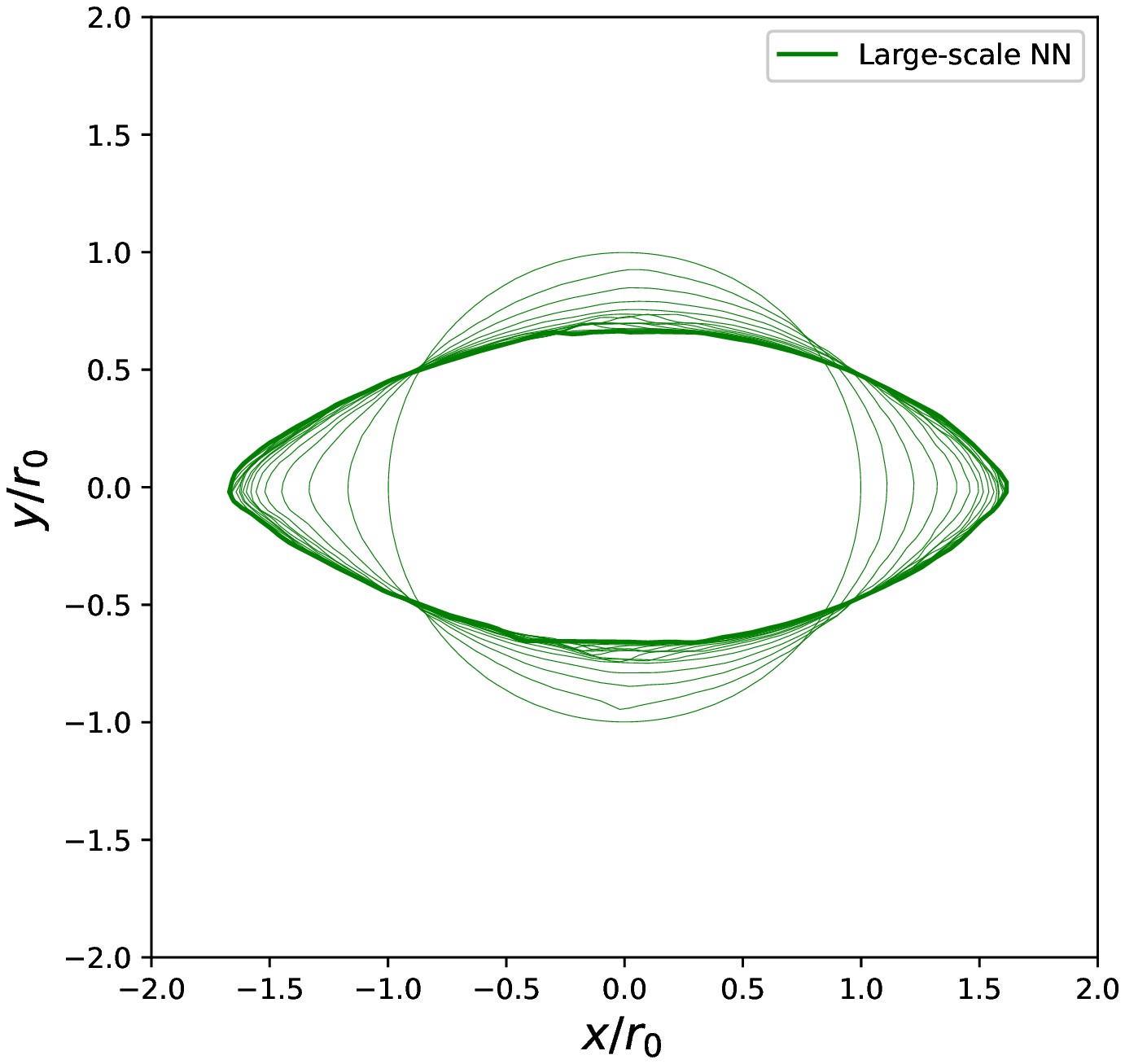}
\caption{}
\end{subfigure}
\caption{The converged shapes at $Re_D=1$ (a) and the intermediate states at every 10th iteration by large-scale NN model (b).}
\label{fig:re1_final}
\end{figure}

\noindent Figures \ref{fig:drag_re1}(a-d) present the drag coefficients over 200 optimisation iterations using \textit{OpenFOAM} solver and three neural network models. Here, the drag coefficient $C_d$ is defined as drag divided by the projected length of the initial cylinder and dynamic head. The same definition is used for all of other experiments in the present paper. As the ground truth, figure \ref{fig:drag_re1}(a) shows the case which uses the \textit{OpenFOAM} solver in the optimisation. The history of drag values, shown in blue, is calculated based on the re-sampled data on the Cartesian grid (i.e. $128^2$). 
For comparison, the drag values obtained from the surface integral in the \textit{OpenFOAM}'s native post-processing code are shown with red markers.
As can be seen in figure \ref{fig:drag_re1}(a), after convergence of the optimisation the total drag drops 6.3\% from 10.43 to 9.78. To further break it down, the inviscid part decreases significantly from 5.20 to 2.50 ($\sim51.8\%$) while the viscous part gradually increases from 5.23 to 7.27 ($\sim31.0\%$). This is associated with the elongation of the shape from a circular cylinder to an "oval", eventually becoming a rugby shape as shown in figure \ref{fig:re1_final}(b). 

From figures \ref{fig:drag_re1}(b-c), one can observe the histories of the drag values are reasonably well predicted by neural network models and agree with the \textit{OpenFOAM} solution in figure \ref{fig:drag_re1}(a). Despite the small scale model exhibiting noticeable oscillations in the optimisation procedure, the medium and large-scale neural network models provide smoother predictions, and the drag of both initial and final shapes agrees well with that from re-sampled data (blue lines) and the one from \textit{OpenFOAM}'s native post-processing code (red symbols). 

Figure \ref{fig:re1_final}(a) depicts the converged shapes
of all four solvers.
The ground truth result using \textit{OpenFOAM} ends up with a rugby shape which achieves a good agreement with the data by \cite{kimkim1995}. The medium and large-scale neural network models collapse and compare favourably with the ground truth result. 
In contrast, the small-scale neural network model's prediction is slightly off which is not surprising as one can observe oscillation and offset of the drag history in figure \ref{fig:drag_re1}(b) as discussed before. A possible reason is that the small scale model has less weights so that the complexity of the flow evolution cannot be fully captured. It is worth noting that the reduced performance of the B{\'e}zier representation in the present work is partly due to the discretization errors when calculating the normal vectors in combination with a reduced number of degrees of freedom.

\begin{figure}

\begin{subfigure}{.45\textwidth}
\centering
\includegraphics[width=\linewidth]{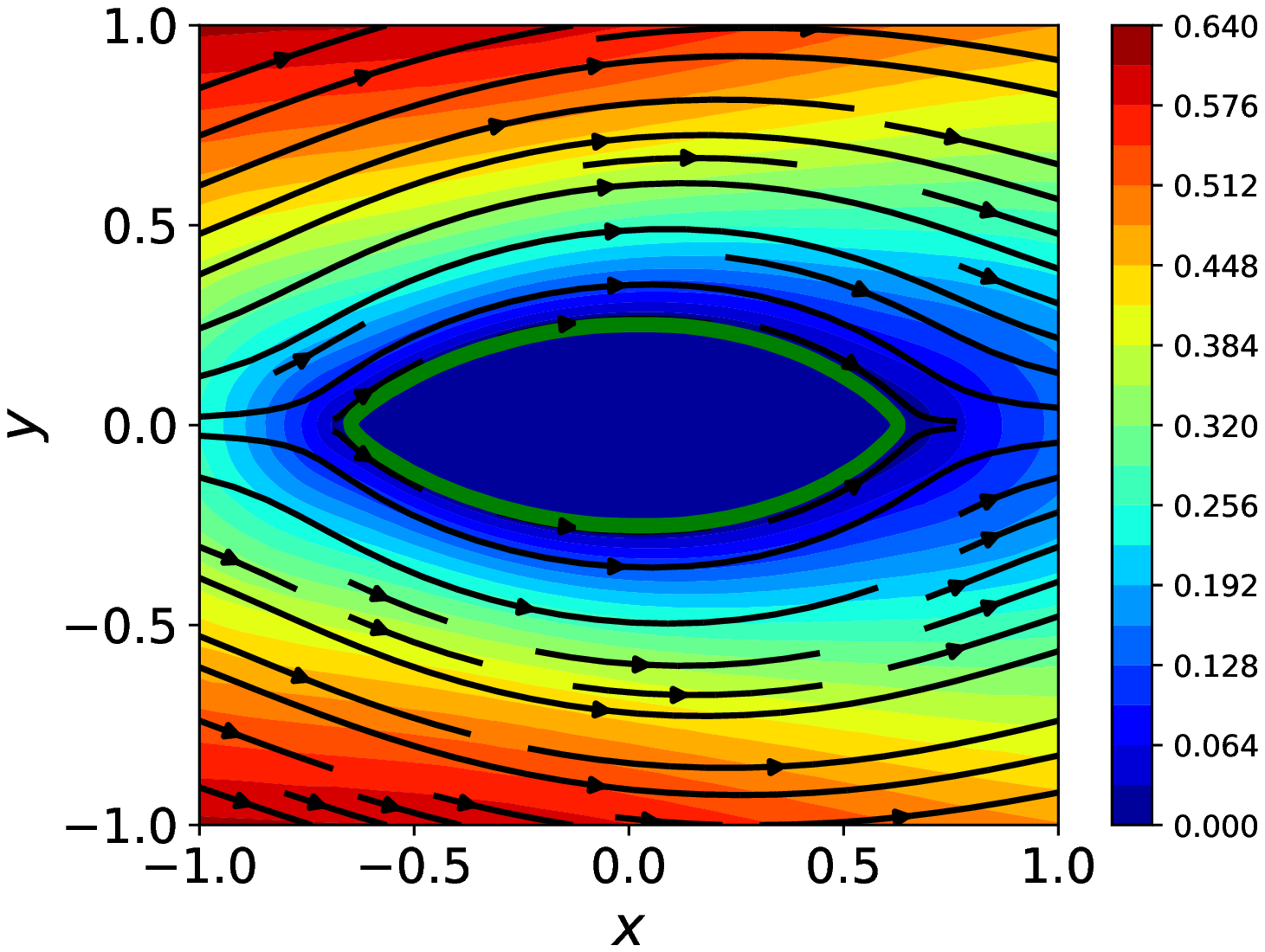}
\caption{OpenFOAM}
\end{subfigure}
\begin{subfigure}{.45\textwidth}
\centering
\includegraphics[width=\linewidth]{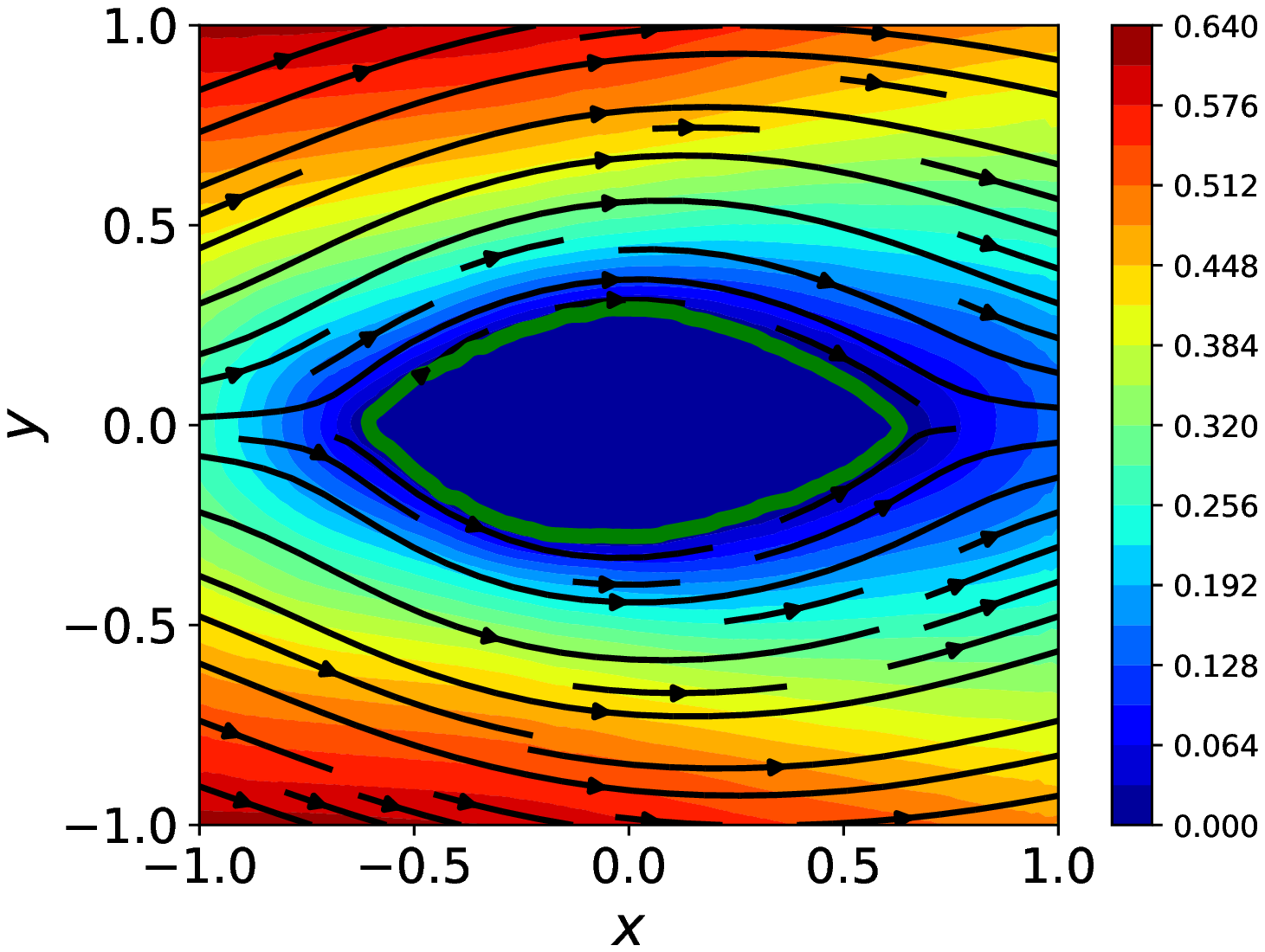}
\caption{Small-scale neural network}
\end{subfigure}

\begin{subfigure}{.45\textwidth}
\centering
\includegraphics[width=\linewidth]{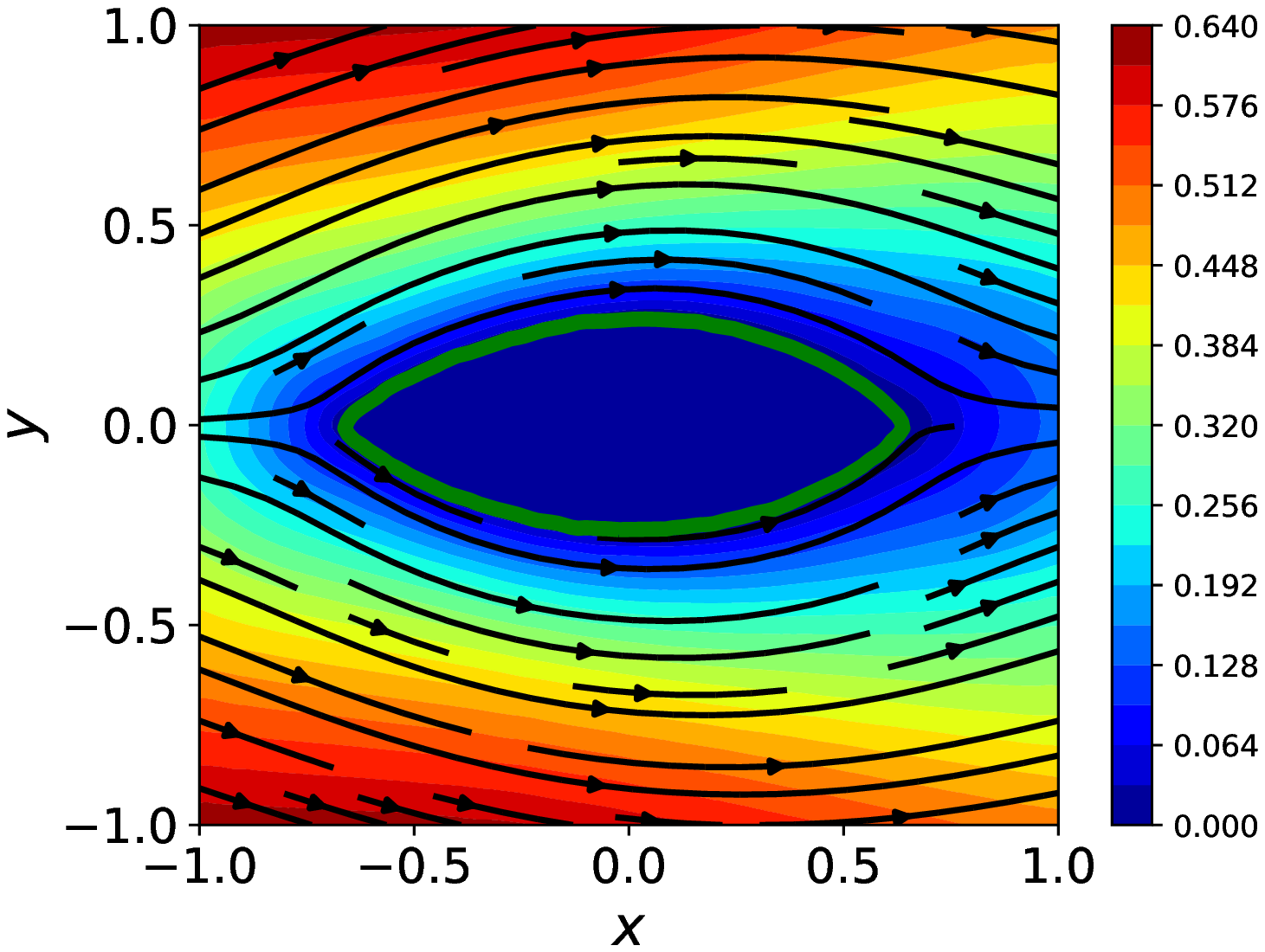}
\caption{Medium-scale neural network}
\end{subfigure}
\begin{subfigure}{.45\textwidth}
\centering
\includegraphics[width=\linewidth]{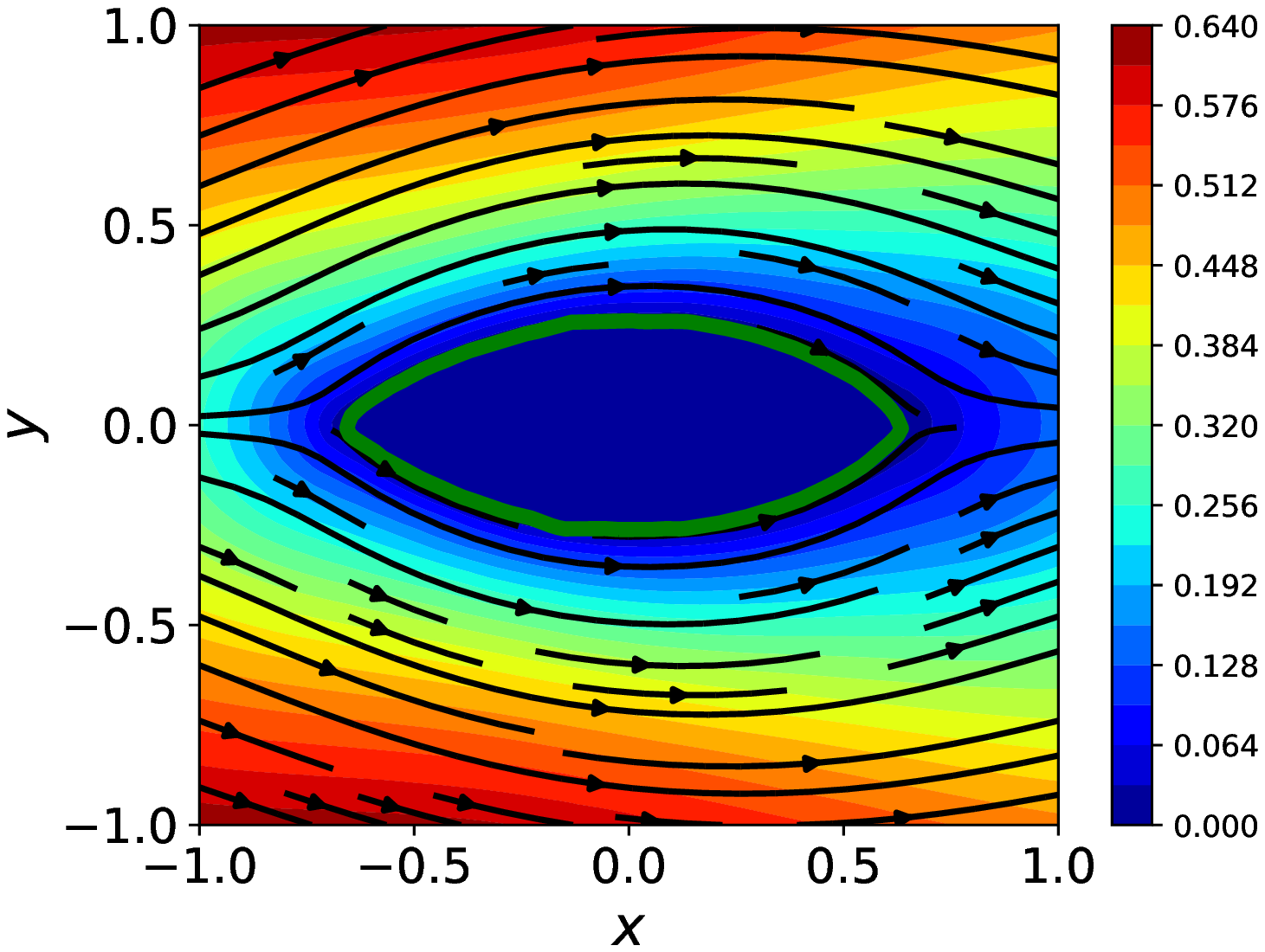}
\caption{Large-scale neural network}
\end{subfigure}

\caption{Streamlines and the x-component velocity fields $u/U_{\infty}$ at $Re_D=1$.}
\label{fig:LSF_re1}
\end{figure}

The x-component velocity fields with streamlines for the optimised shapes are shown in figure \ref{fig:LSF_re1}. The flowfields and the patterns of streamlines in all the three cases with neural networks show no separation, which is consistent with the ground truth result in figure \ref{fig:LSF_re1}(a).
Considering the final shape obtained using the three neural network surrogates, the medium- and large-scale models give satisfactory results that are close to the \textit{OpenFOAM} result.

\subsection{Optimisation experiment at $Re_D=40$}

\begin{figure}

\begin{subfigure}{.45\textwidth}
\centering
\includegraphics[width=\linewidth]{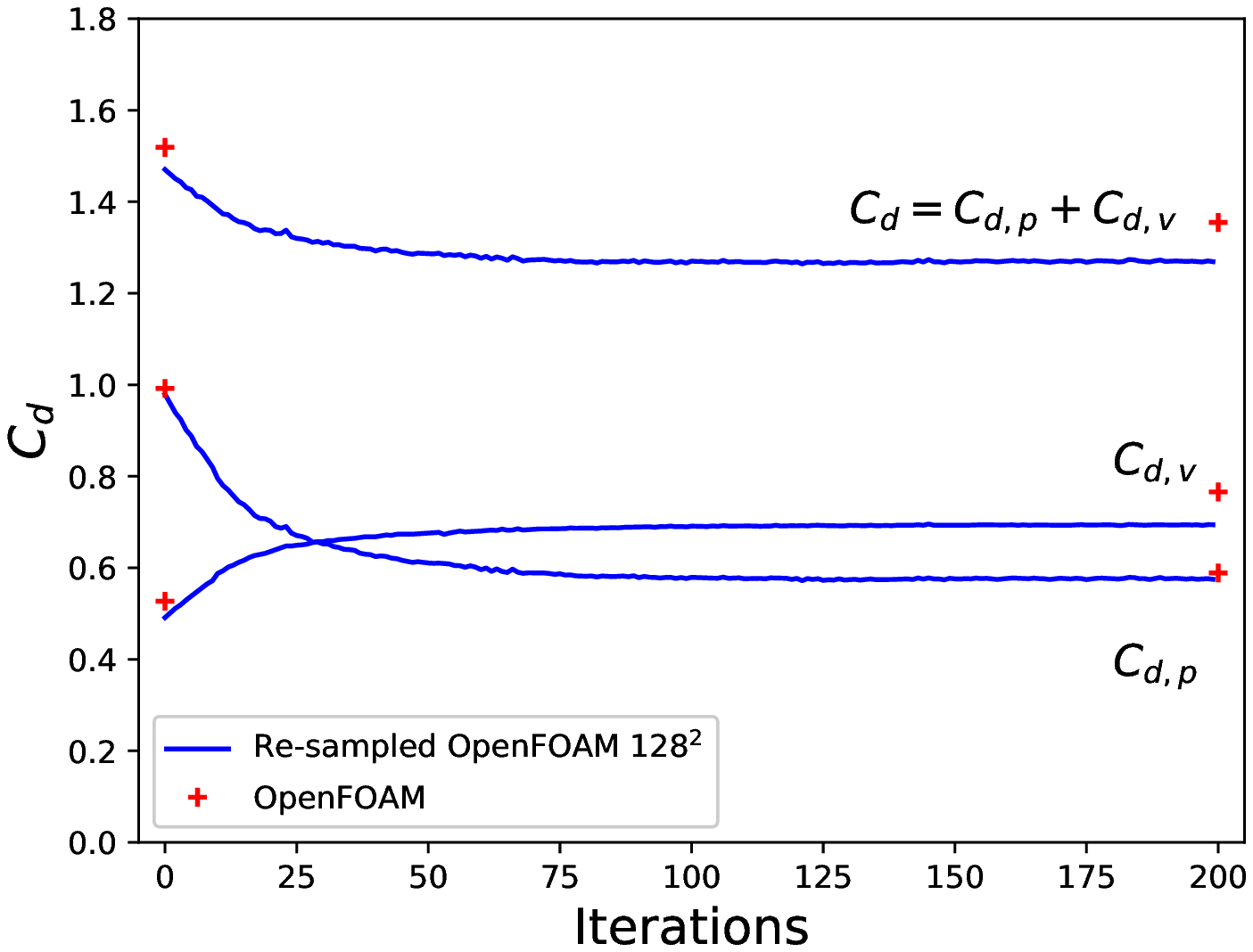}
\caption{OpenFOAM}
\end{subfigure}
\begin{subfigure}{.45\textwidth}
\centering
\includegraphics[width=\linewidth]{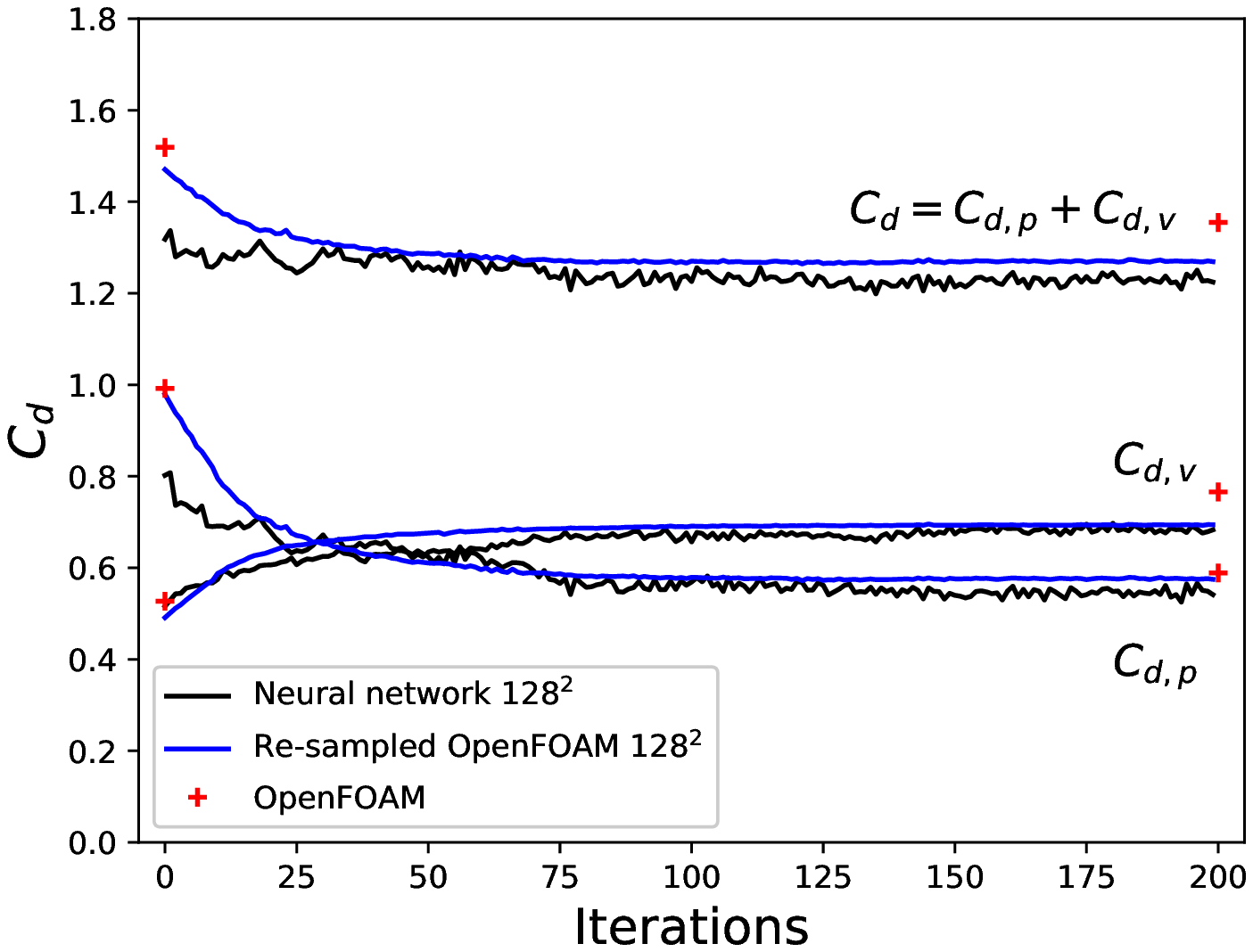}
\caption{Small-scale neural network}
\end{subfigure}

\begin{subfigure}{.45\textwidth}
\centering
\includegraphics[width=\linewidth]{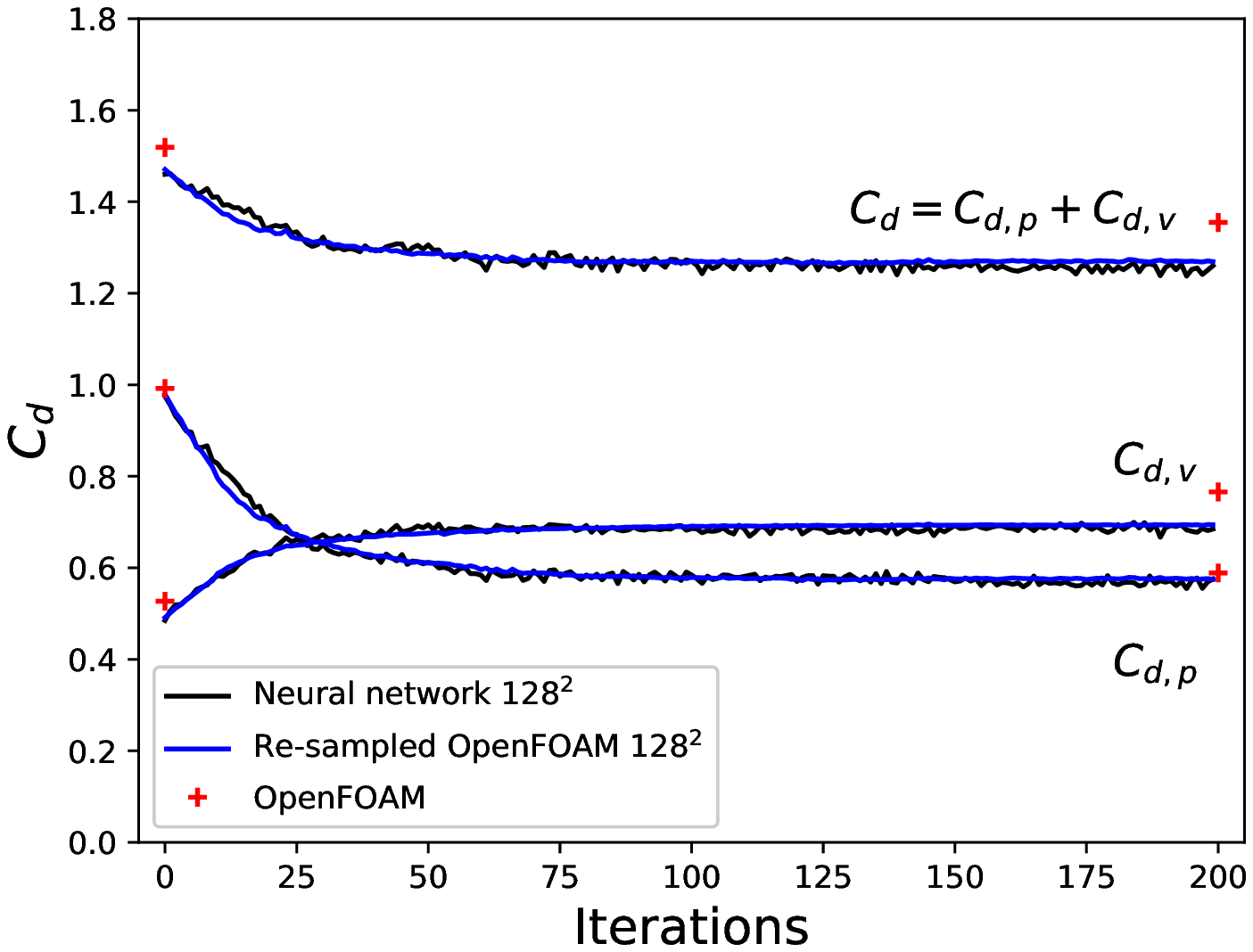}
\caption{Medium-scale neural network}
\end{subfigure}
\begin{subfigure}{.45\textwidth}
\centering
\includegraphics[width=\linewidth]{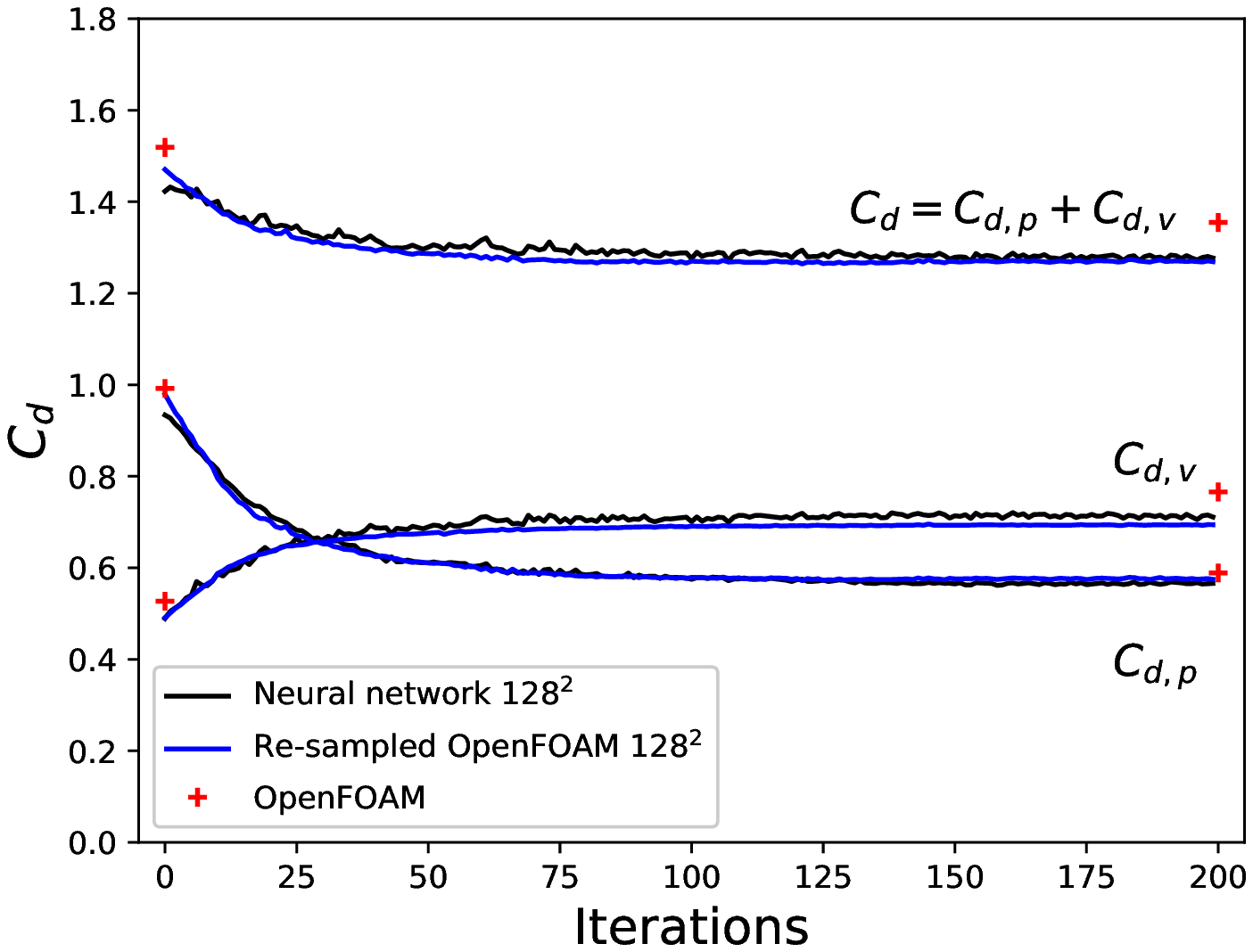}
\caption{Large-scale neural network}
\end{subfigure}

\caption{Optimisation histories at $Re_D=40$. The black solid lines denote the results using neural network models trained with ``Dataset-40'' and the blue solid lines denote the results from \textit{OpenFOAM}. Results calculated with the re-sampled flowfields on the $128\times128$ Cartesian grid are denoted by $128^2$. The red cross symbols represent the \textit{OpenFOAM}'s results obtained with its native postprocessing tool.}
\label{fig:drag_re40}
\end{figure}

\begin{figure}
\begin{subfigure}{.45\textwidth}
\centering
\includegraphics[width=\linewidth]{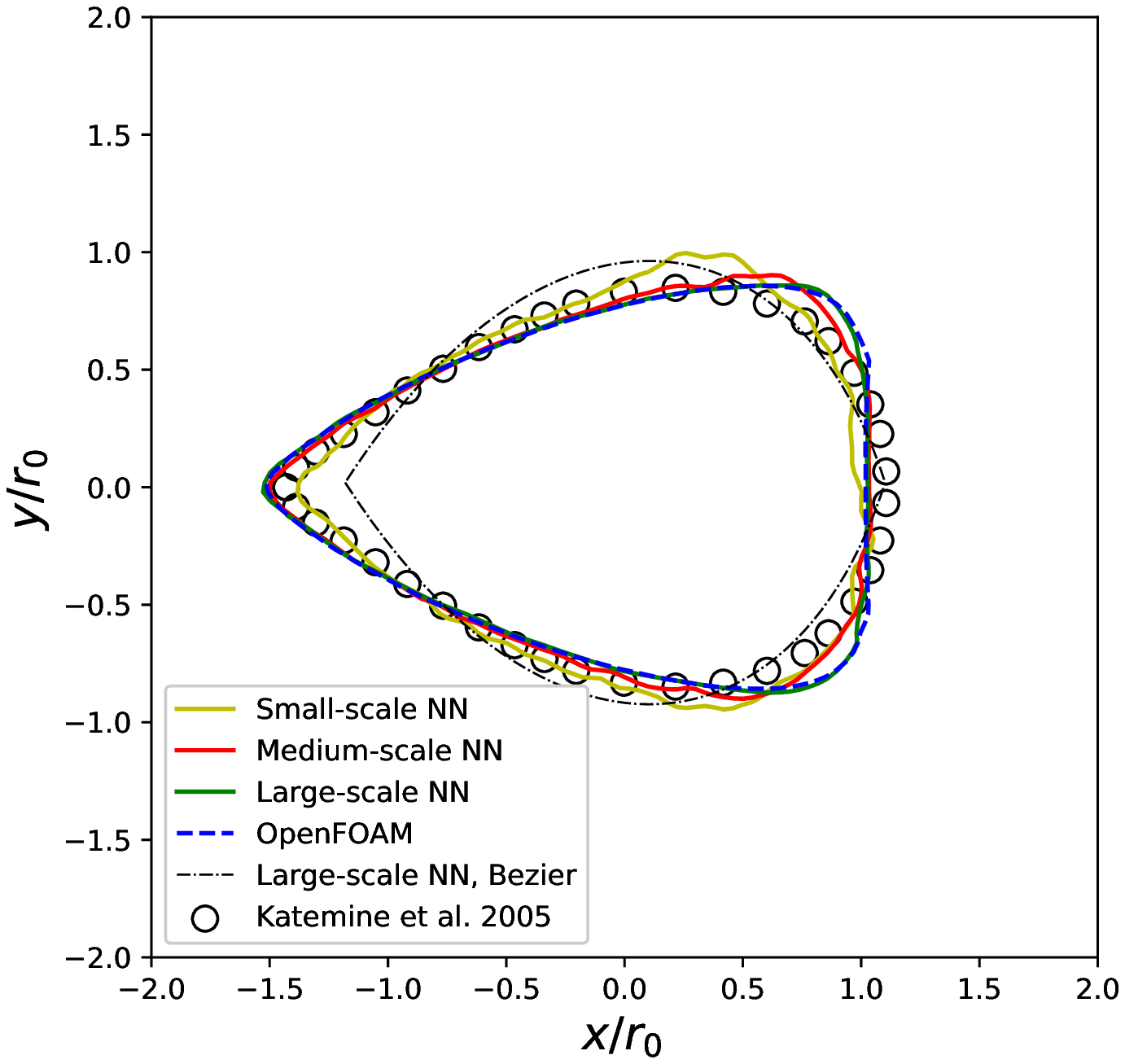}
\caption{}
\end{subfigure}
\begin{subfigure}{.45\textwidth}
\centering
\includegraphics[width=\linewidth]{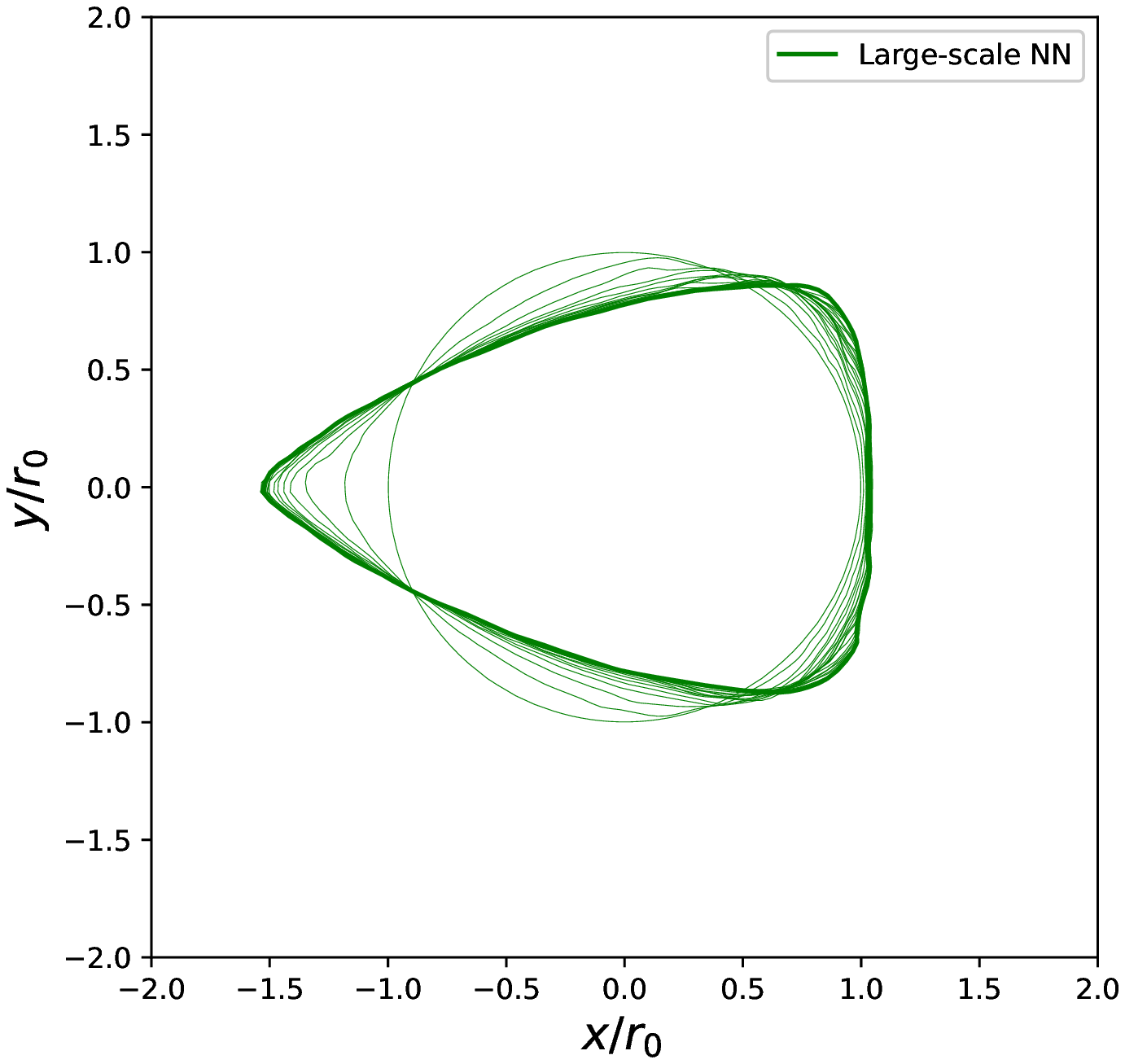}
\caption{}
\end{subfigure}
\caption{The converged shapes at $Re_D=40$ (a) and the intermediate states at every 10th iteration by large-scale NN model (b).}
\label{fig:re40_final}
\end{figure}

\begin{figure}

\begin{subfigure}{.45\textwidth}
\centering
\includegraphics[width=\linewidth]{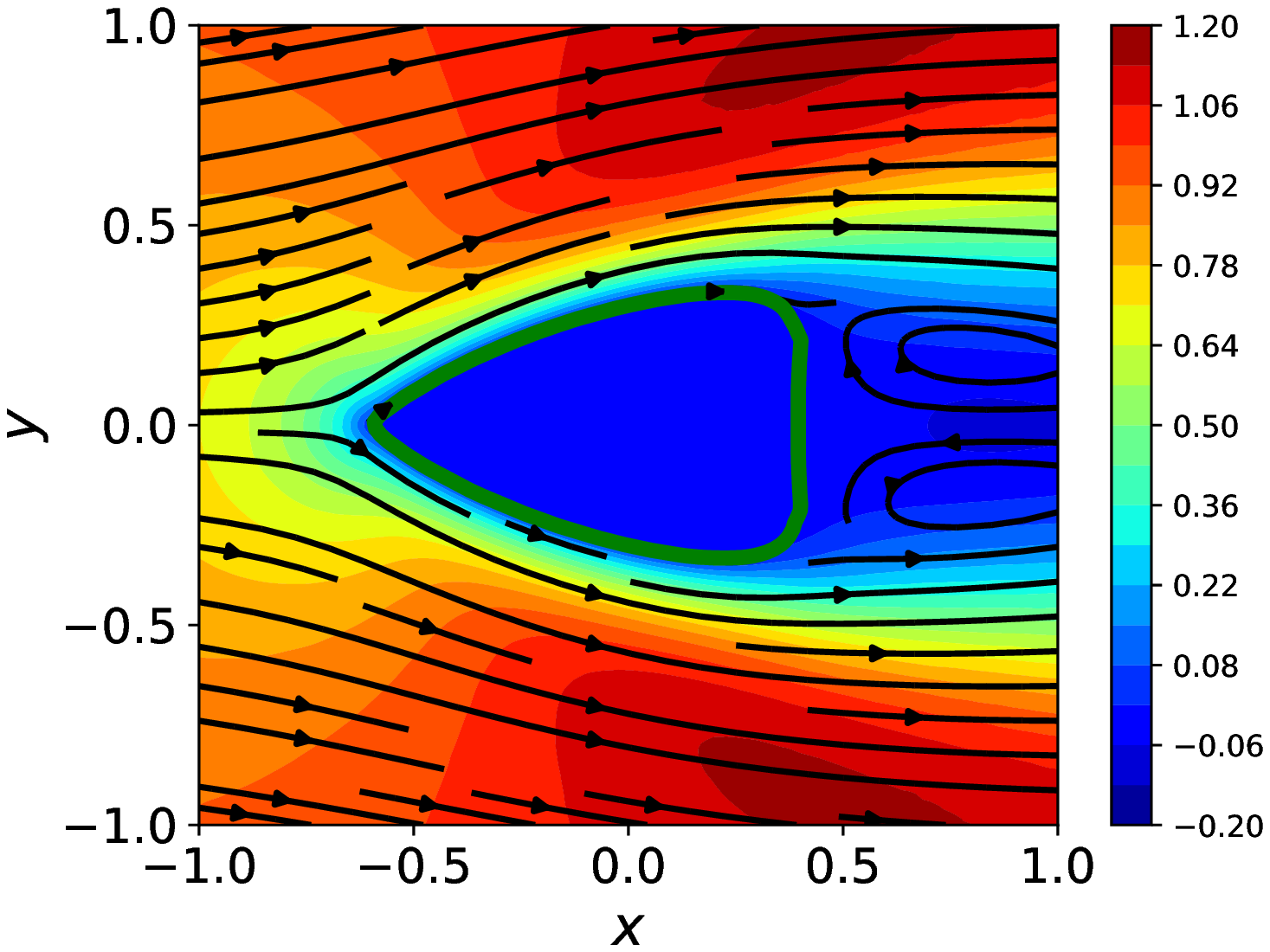}
\caption{OpenFOAM}
\end{subfigure}
\begin{subfigure}{.45\textwidth}
\centering
\includegraphics[width=\linewidth]{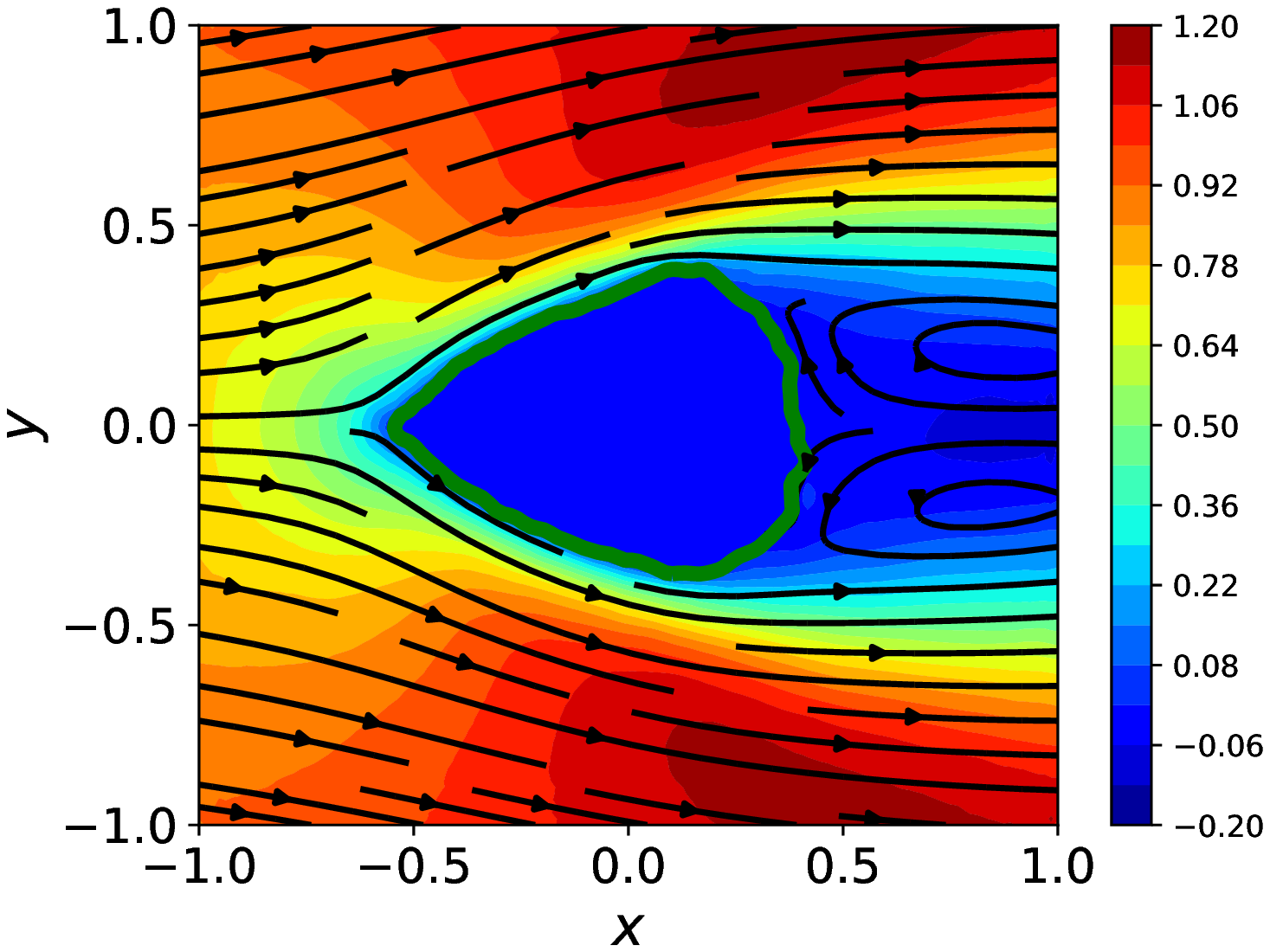}
\caption{Small-scale neural network}
\end{subfigure}

\begin{subfigure}{.45\textwidth}
\centering
\includegraphics[width=\linewidth]{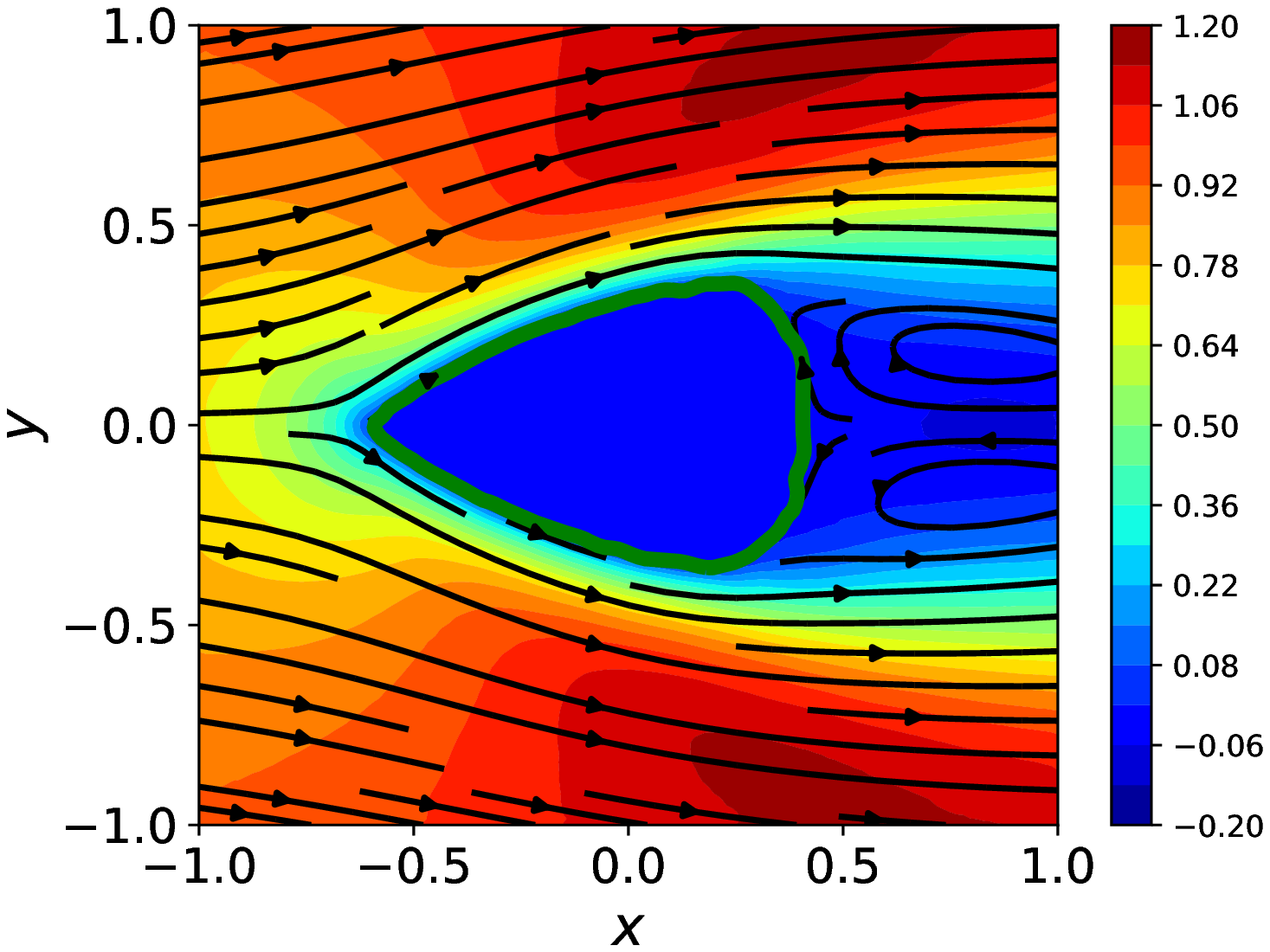}
\caption{Medium-scale neural network}
\end{subfigure}
\begin{subfigure}{.45\textwidth}
\centering
\includegraphics[width=\linewidth]{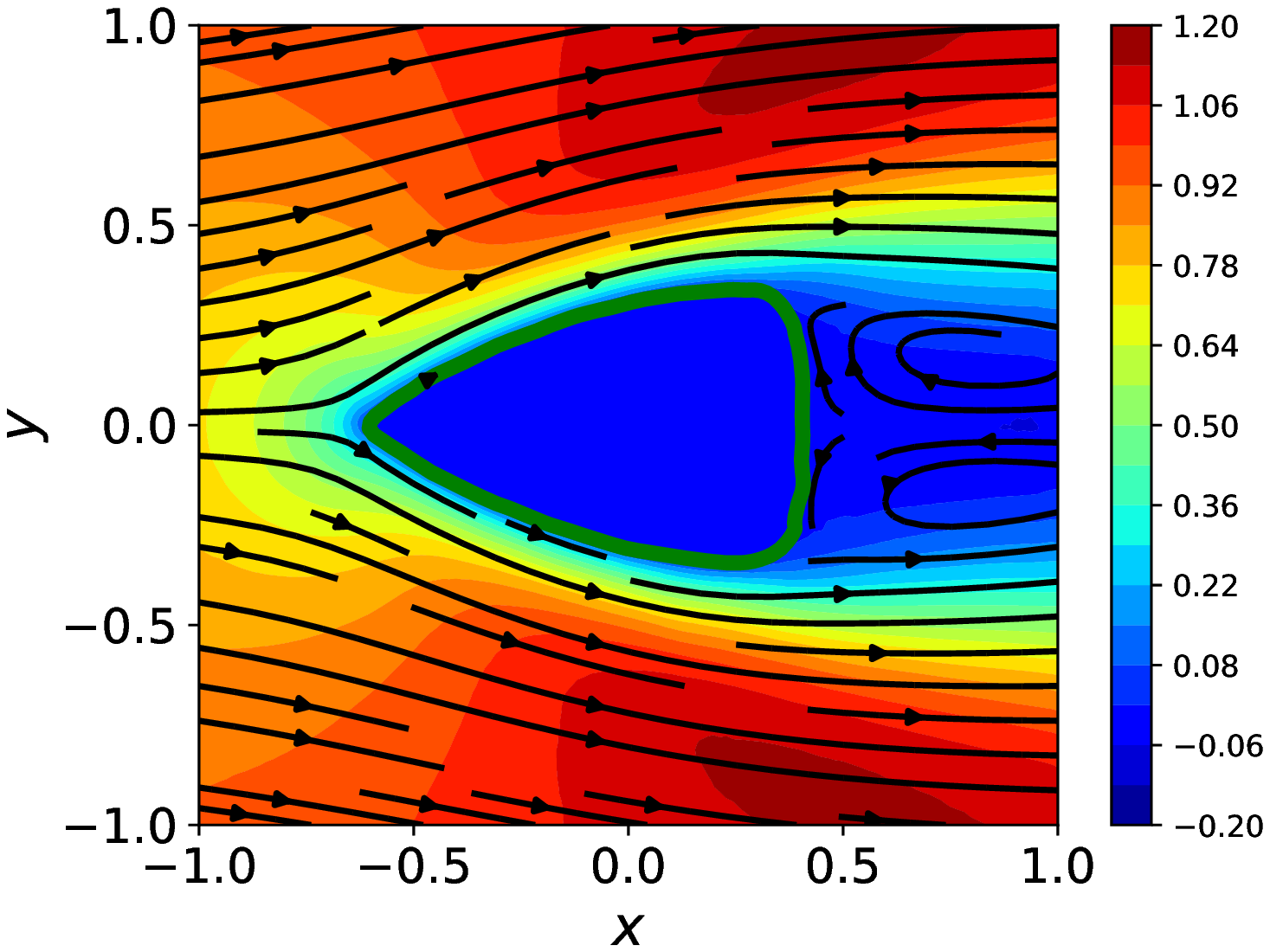}
\caption{Large-scale neural network}
\end{subfigure}

\caption{Streamlines and the x-component velocity fields $u/U_{\infty}$ at $Re_D=40$ obtained with different solvers, i.e. \textit{OpenFOAM}, and three neural network models trained with ``Dataset-40''.}
\label{fig:LSF_re40}
\end{figure}

\noindent As the Reynolds number increases past the critical Reynolds number $Re_D\approx47$, the circular cylinder flow configuration loses its symmetry and becomes unstable, which is known as the Karman vortex street. We consider optimisations for the flow regime at $Re_D=40$ which is of particular interest because it exhibits a steady-state solution, yet is close to the critical Reynolds number. The steady separation bubbles behind the profile further compound the learning task and the optimisation, making it a good test case for the proposed method.

The ground truth optimisation result using \textit{OpenFOAM} is shown in figure \ref{fig:drag_re40}(a). The shape is initialised with a circular cylinder and is optimised to minimise drag over 200 iterations. As a result, the total drag, processed on the Cartesian grid, drops from 1.470 to 1.269 ($\sim13.7\%$ reduction). Associated with the elongation of the shape, the inviscid drag decreases $41.3\%$ while the viscous drag increases $41.3\%$. The initial and the final results of the \textit{OpenFOAM}'s native post-processing are shown in red, indicating good agreement. Figures \ref{fig:drag_re40}(b-d) present the drag histories over 200 optimisation iterations with three neural network models that are trained with ``Dataset-40''. Although larger oscillations are found in the drag history of the small scale model, the medium and large scale models predict smoother drag history and compare well with the ground truth data using \textit{OpenFOAM}. 

The final converged shapes are compared to a reference result \citep{katamine2005} in figure \ref{fig:re40_final}(a). The evolution of intermediate shapes from the initial circular cylinder towards the final shape is shown in figure \ref{fig:re40_final}(b). The upwind side forms a sharp leading edge while the downwind side of the profile develops into a blunt trailing edge. Compared to the reference data \citep{katamine2005} and the result using B{\'e}zier-curve based method, the use of level-set based method leads to a slightly flatter trailing edge, probably because more degrees of freedom for the shape representation are considered in level-set based method.

Further looking at the details of shapes in figure \ref{fig:LSF_re40}, it can be seen that the 
more weights the neural network model contains,
the closer it compares to the ground truth result using \textit{OpenFOAM}. The large scale model which has the largest weight count is able to resolve the fine feature of the flat trailing edge as shown in figure \ref{fig:LSF_re40}(d). In contrast, in figure \ref{fig:LSF_re40}(b), the small scale model does not capture that and even the the surface of the profile exhibits pronounced roughness. Nonetheless, all the three DNN models predict similar flow patterns compared to the ground truth result depicted with streamlines, which are characterised with re-circulation regions downstream of the profiles.

It should be mentioned that the optimised shape at $Re_D=40$ by \citet{kimkim1995} differs from the one in the present study and the one by \citet{katamine2005}. In the former \citep{kimkim1995}, the optimised profile converges at an elongated slender shape with an even smaller drag force. 
Most likely, this is caused by that an additional wedge angle constraint is imposed at both leading and trailing edge, which is not adopted in our work and \citet{katamine2005}. 
As we focus on deep learning surrogates in the present study, we believe the topic of including additional constraints will be an interesting avenue for future work.
In the comparison to the ground truth from \textit{OpenFOAM}, the current results are deemed to be in very good agreement.

\subsection{Shape optimisations for an enlarged solution space}
\noindent The generalising capabilities of neural networks are a challenging topic \citep{ling2016reynolds}. 
To evaluate their flexibility in our context, we target shape optimisations
in the continuous range of Reynolds numbers from $Re_D=1$ to 40, over the course of which
the flow patterns change significantly \citep{tritton_1959, sen_mittal_biswas_2009}. 
Hence, in order to succeed, a neural network not only has to encode change of the solutions w.r.t. immersed 
shape but also the changing physics of the different Reynolds numbers.
In this section, we conduct four tests at $Re_D=1$, 5, 10, and 40 with the ranged model
in order to quantitatively assess the its ability to make accurate flowfield predictions 
over the chosen change of Reynolds numbers.
The corresponding \textit{OpenFOAM} runs are used as ground truth for comparisons.

\begin{figure}

\begin{subfigure}{.45\textwidth}
\centering
\includegraphics[width=\linewidth]{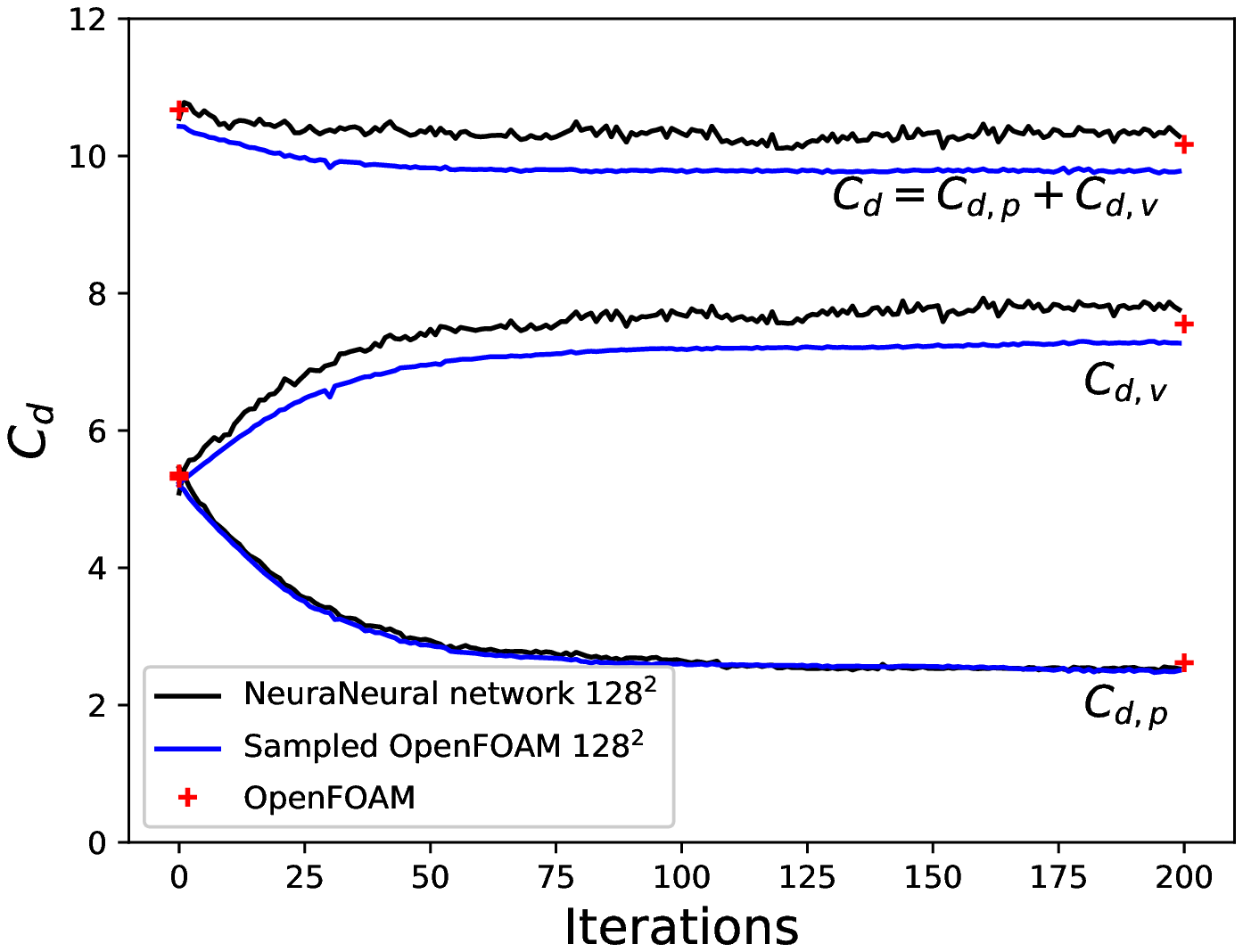}
\caption{$Re_D=1$}
\end{subfigure}
\begin{subfigure}{.45\textwidth}
\centering
\includegraphics[width=\linewidth]{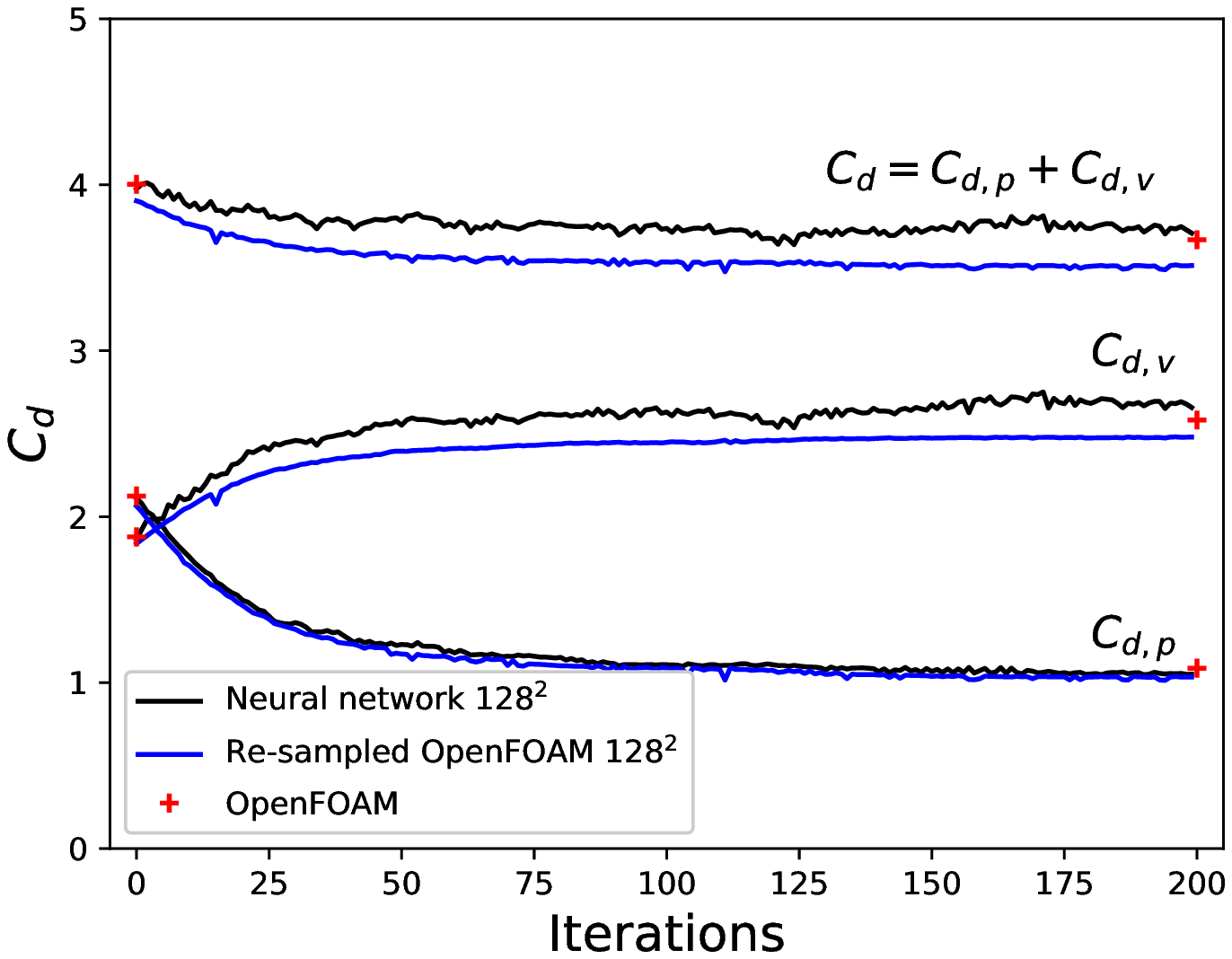}
\caption{$Re_D=5$}
\end{subfigure}

\begin{subfigure}{.45\textwidth}
\centering
\includegraphics[width=\linewidth]{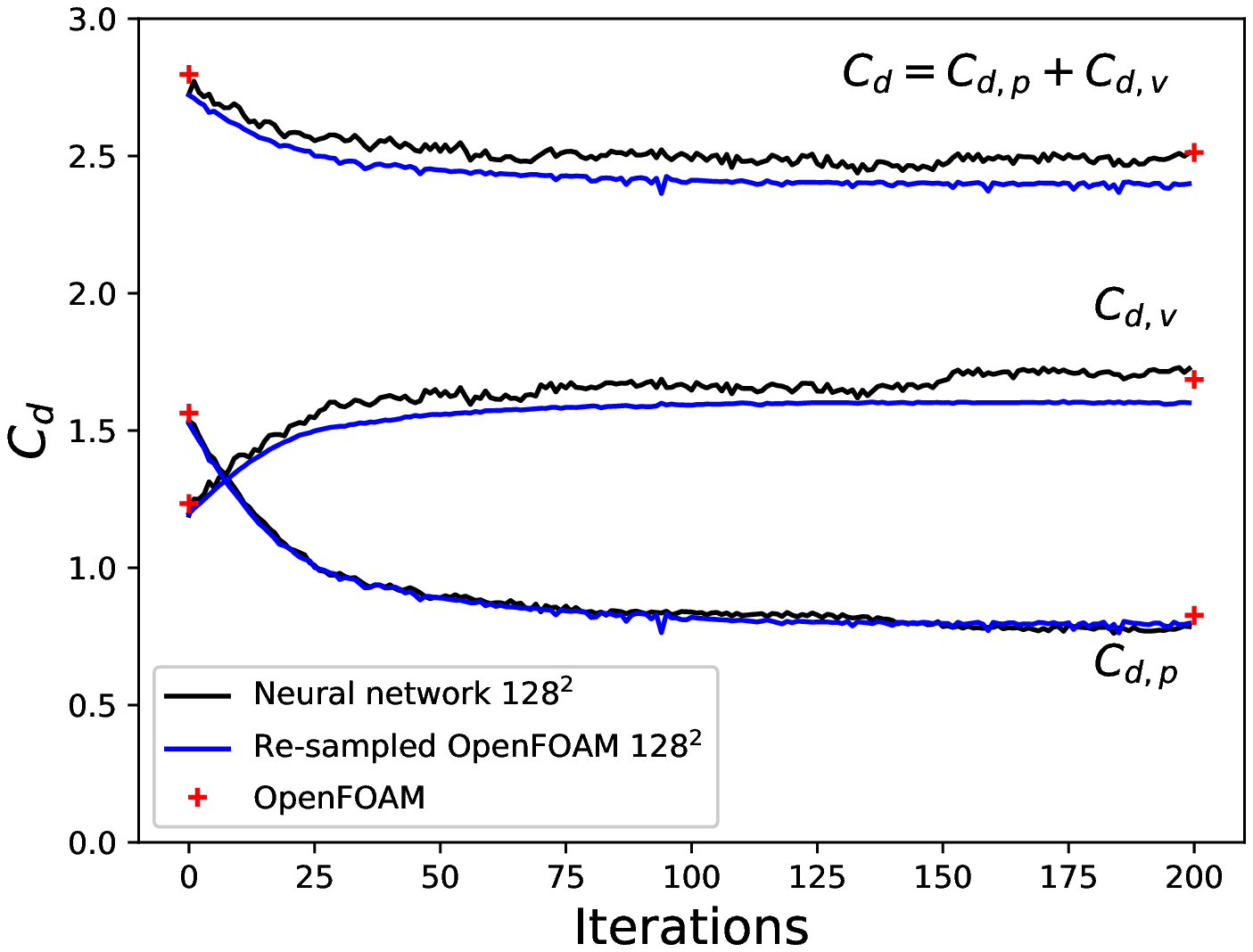}
\caption{$Re_D=10$}
\end{subfigure}
\begin{subfigure}{.45\textwidth}
\centering
\includegraphics[width=\linewidth]{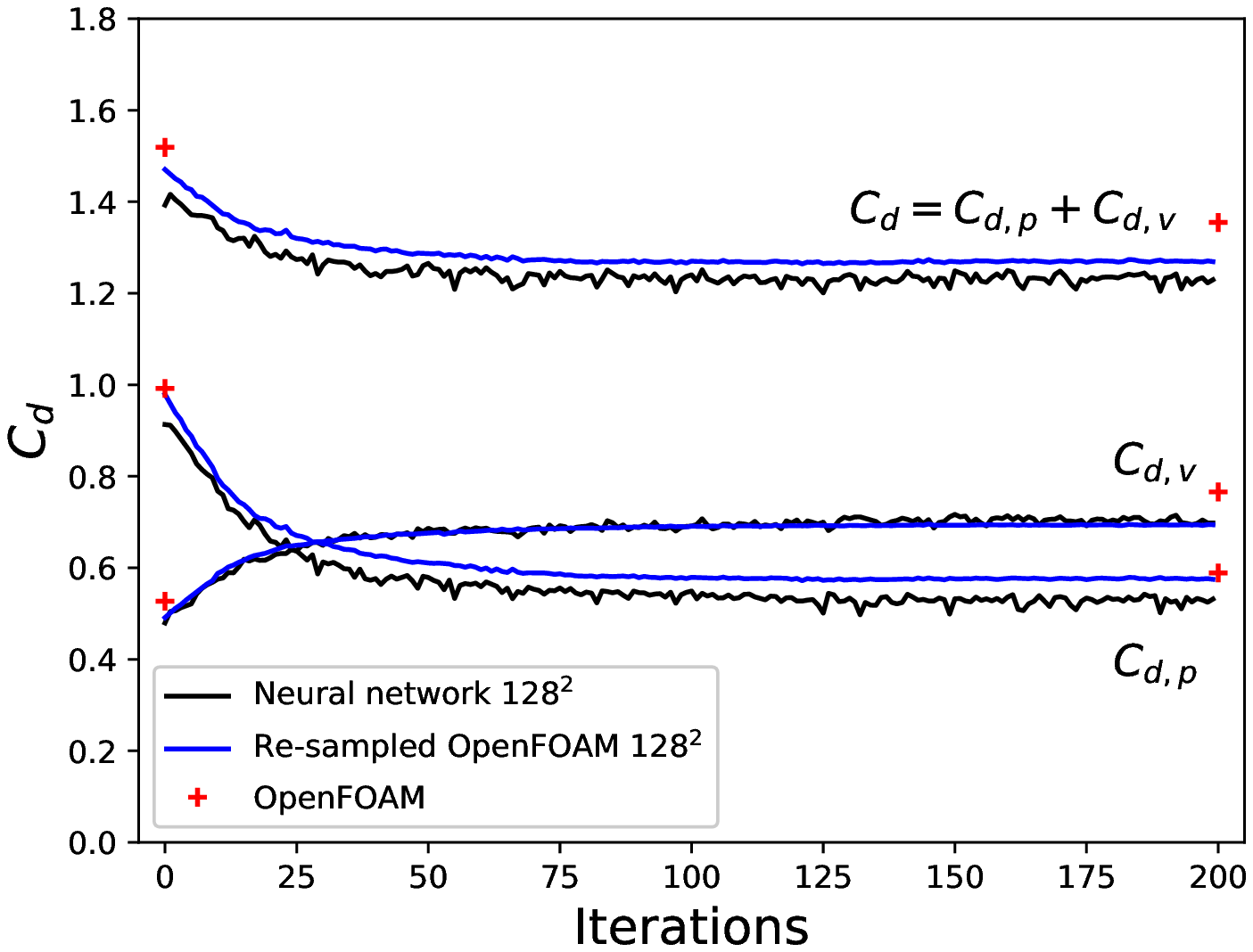}
\caption{$Re_D=40$}
\end{subfigure}
\caption{Optimisation history for the four cases at $Re_D=1$, 5, 10, and 40. The black solid lines denote the results using neural network models (i.e. the ranged model) and the blue solid lines denote the results from \textit{OpenFOAM}. Results calculated with the re-sampled flowfields on the $128\times128$ Cartesian grid are denoted by $128^2$. The red cross symbols represent the \textit{OpenFOAM} results obtained with its native postprocessing tool.}
\label{fig:drag_history_general}
\end{figure}

The optimisation histories for the four cases are plotted in figures \ref{fig:drag_history_general}(a-d). Despite some oscillations, the predicted drag values as well as the inviscid and viscous parts agree well with the ground truth values from \textit{OpenFOAM}. The total drag force as objective function has been reduced and reaches a stable state in each case. The performance of the ranged model at $Re_D=40$ is reasonably good, although 
it is slightly outperformed by the specialized NN model trained with ``Dataset-40''. 

\begin{figure}
\begin{subfigure}{.45\textwidth}
\centering
\includegraphics[width=\linewidth]{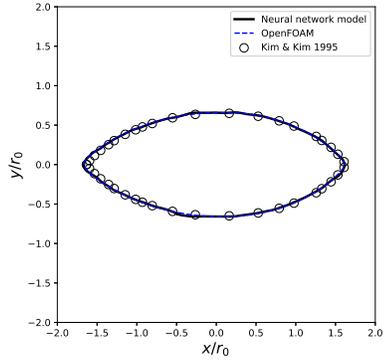}
\caption{$Re_D=1$}
\end{subfigure}
\begin{subfigure}{.45\textwidth}
\centering
\includegraphics[width=\linewidth]{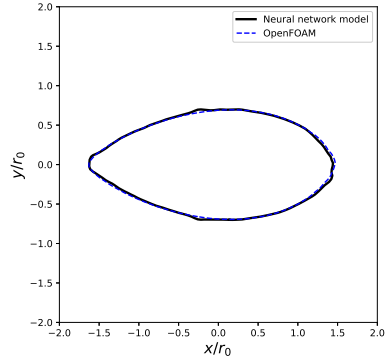}
\caption{$Re_D=5$}
\end{subfigure}

\begin{subfigure}{.45\textwidth}
\centering
\includegraphics[width=\linewidth]{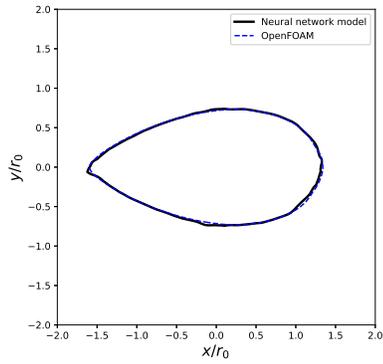}
\caption{$Re_D=10$}
\end{subfigure}
\begin{subfigure}{.45\textwidth}
\centering
\includegraphics[width=\linewidth]{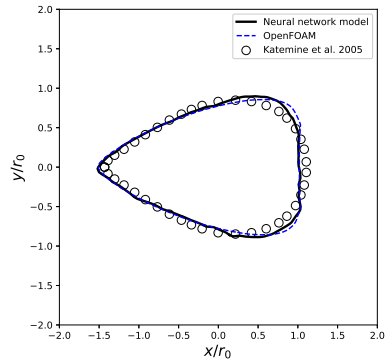}
\caption{$Re_D=40$}
\end{subfigure}
\caption{Shapes after optimisation at $Re_D=1$, 5, 10, and 40. The black solid lines denote the results using neural network models (i.e. the ranged model), the blue dashed lines denote the results from \textit{OpenFOAM} and the symbols denote the corresponding reference data.}
\label{fig:shape_final_general}
\end{figure}

\begin{figure}

\begin{subfigure}{.45\textwidth}
\centering
\includegraphics[width=\linewidth]{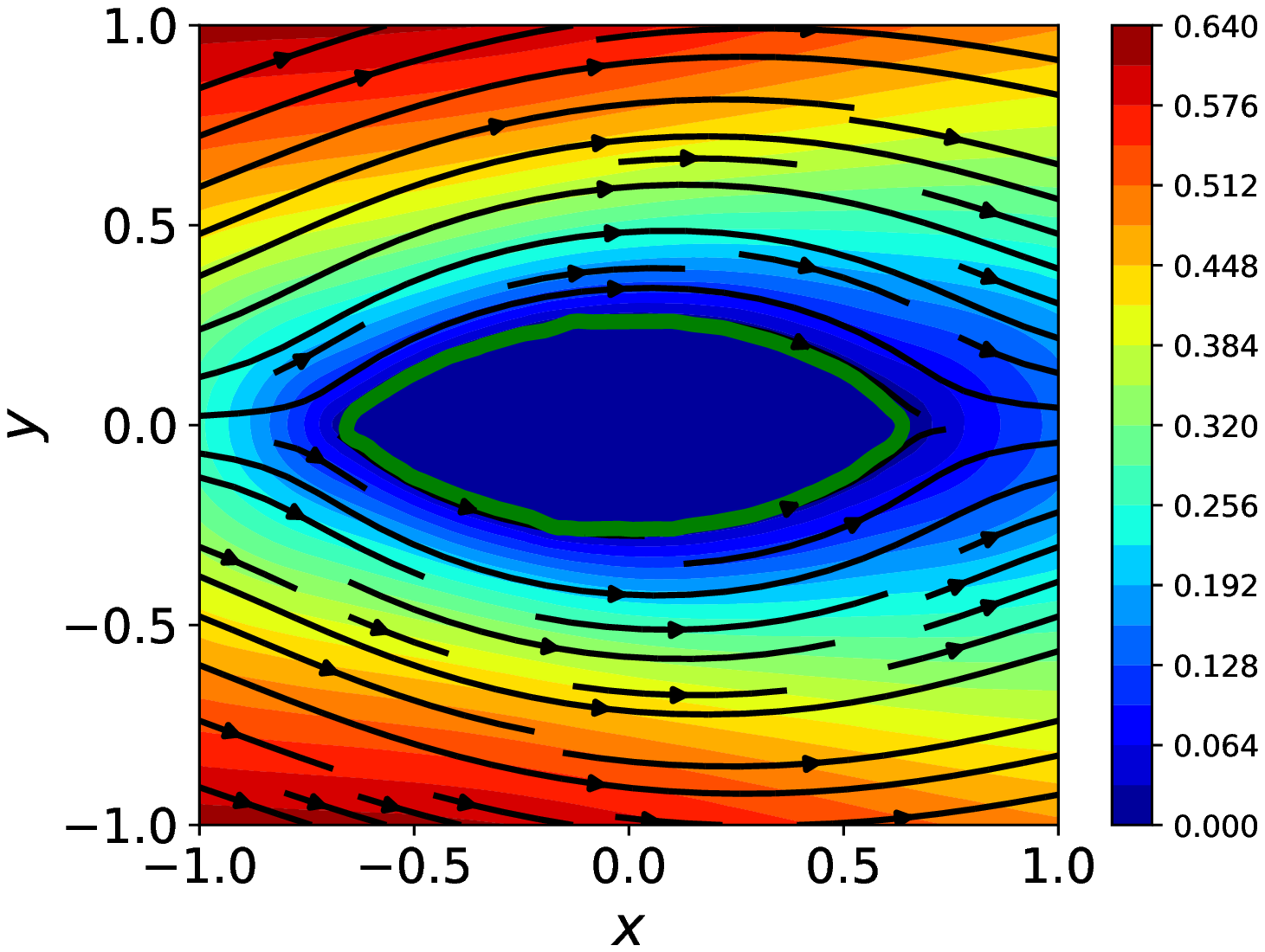}
\caption{$Re_D=1$}
\end{subfigure}
\begin{subfigure}{.45\textwidth}
\centering
\includegraphics[width=\linewidth]{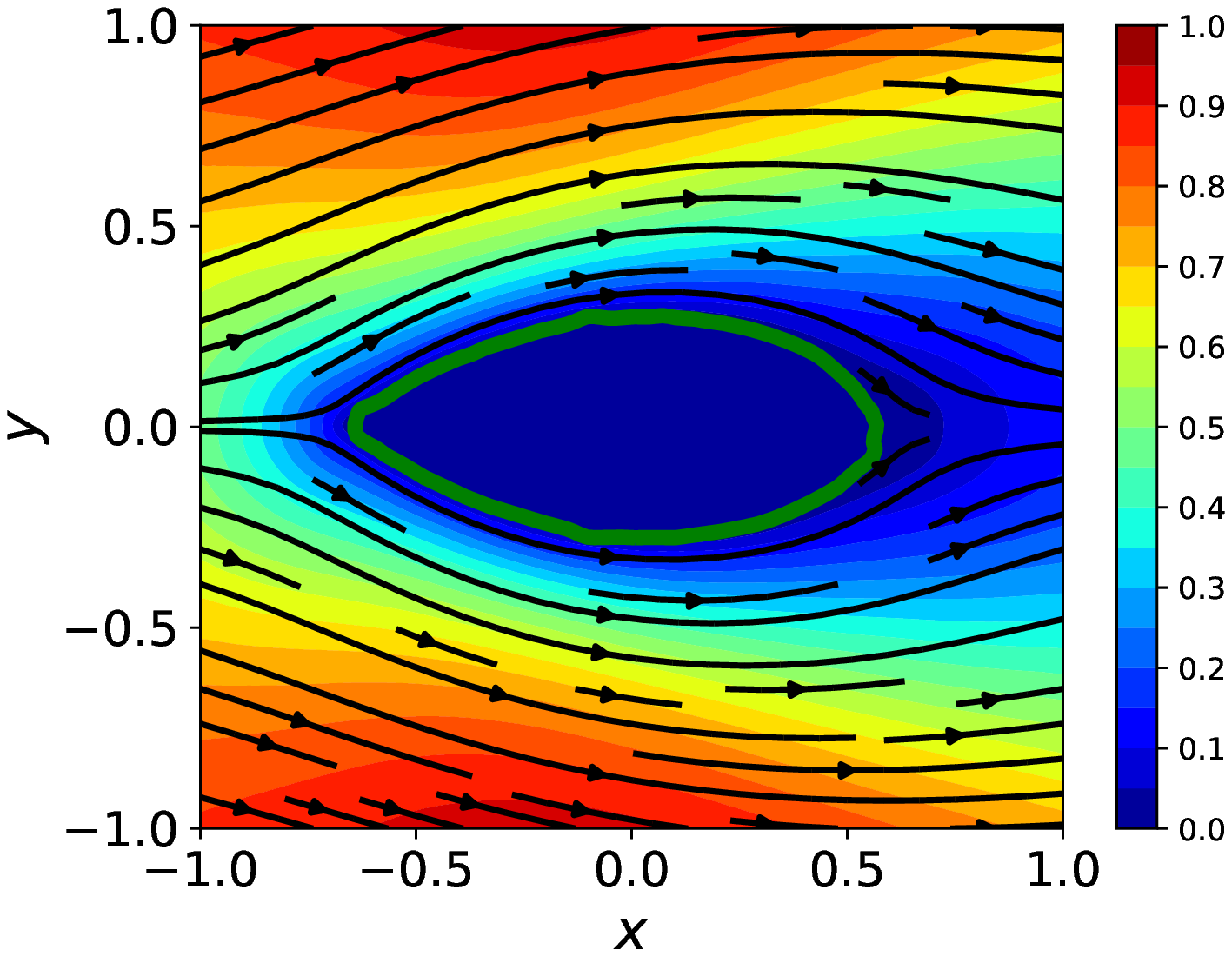}
\caption{$Re_D=5$}
\end{subfigure}

\begin{subfigure}{.45\textwidth}
\centering
\includegraphics[width=\linewidth]{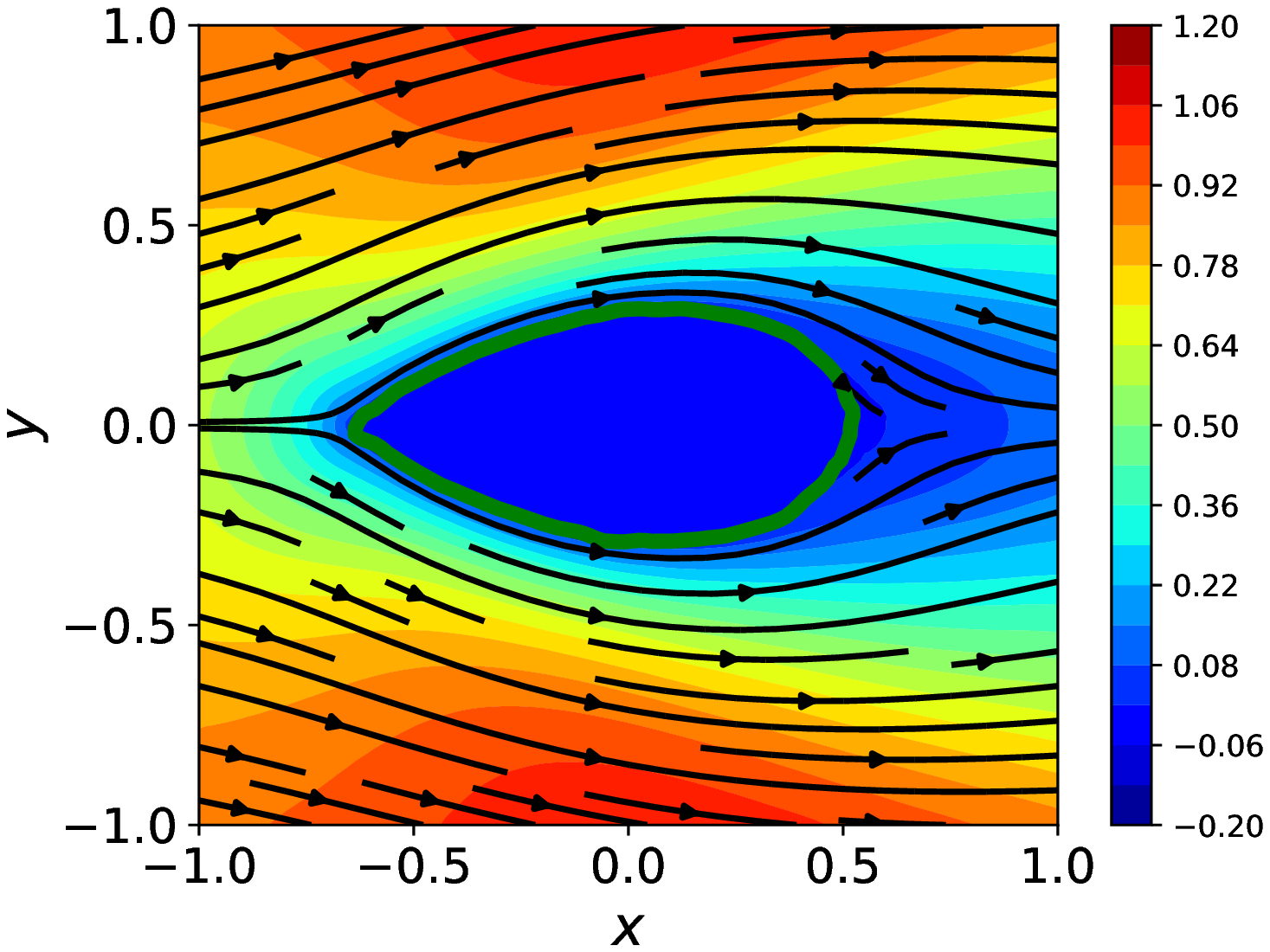}
\caption{$Re_D=10$}
\end{subfigure}
\begin{subfigure}{.45\textwidth}
\centering
\includegraphics[width=\linewidth]{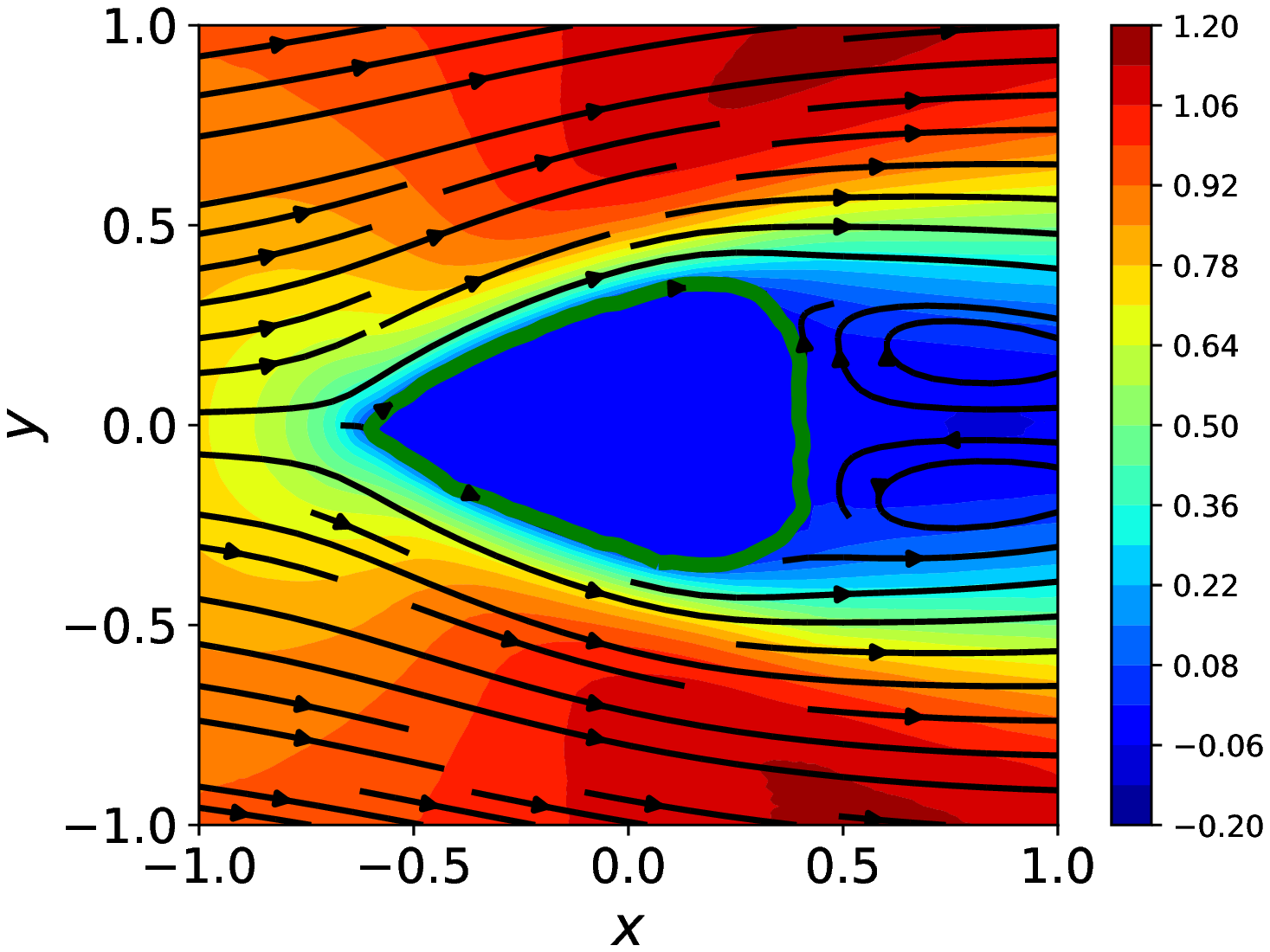}
\caption{$Re_D=40$}
\end{subfigure}

\caption{Streamlines and the x-component velocity fields $u/U_{\infty}$ at $Re_D=1$, 5, 10, and 40 using the ranged model.}
\label{fig:LSF_reGeneral}
\end{figure}

In line with the previous runs, the overall trend of optimisation for the four cases shows that the viscous drag increases while the inviscid part decreases as shown in figures \ref{fig:drag_history_general}(a-d), which is associated with an elongation of the profile and the formation of the sharp leading edge. The final shapes after optimisation for four Reynolds numbers are summarised in figure \ref{fig:shape_final_general}. For the four cases, the leading eventually develops a sharp leading edge, while the trailing edge shows difference. At $Re_D=1$ and 5, the profiles converge with sharp trailing edges as depicted in figures \ref{fig:shape_final_general}(a) and \ref{fig:shape_final_general}(b). The corresponding flowfields also show no separations in figures \ref{fig:LSF_reGeneral}(a) and \ref{fig:LSF_reGeneral}(b).

As shown in figure \ref{fig:shape_final_general}(c) at $Re_D=10$ and figure \ref{fig:shape_final_general}(d) at $Re_D=40$, blunt trailing edges become the final shapes and the profile at $Re_D=10$ is more slender than that for $Re_D=40$. The higher Reynolds number leads to a flattened trailing edge,  associated with the occurrence of the recirculation region shown in figures \ref{fig:LSF_reGeneral}(c) and \ref{fig:LSF_reGeneral}(d), and the gradient of the objective function becoming relatively week in these regions. In terms of accuracy, the converged shapes at $Re_D=1$, 5, and 10 compare favourably with the results with \textit{OpenFOAM}. Compared to ground truth shapes, only the final profile at $Re_D=40$ predicted by the ranged model shows slight deviations near trailing edge. 
Thus, given the non-trivial changes of flow behavior across the targeted range or Reynolds numbers, the neural network yields a robust and consistent performance.

\subsection{Performance}
\begin{table}
  \begin{center}
\def~{\hphantom{0}}
  \begin{tabular}{lll}
      \toprule
      Solver    & Wall time & Platform \\[3pt]
      \midrule 
      \textit{OpenFOAM}    & 16.3 hr & CPU only, 9 cores \\
      Small-scale DNN      & 97 sec	 & CPU, 1 core \& GPU \\
      Medium-scale DNN     & 106 sec &  CPU, 1 core \& GPU \\
      Large-scale DNN      & 196 sec & CPU, 1 core \& GPU \\
      \bottomrule
  \end{tabular}
  \caption{Run times for 200 optimisation iterations at $Re_D=1$. 
  }
  \label{tab:performance}
  \end{center}
\end{table}

\noindent The performance of trained deep neural network models is one of the central factors motivating their use. We evaluate our models in a standard workstation with 12 cores, i.e. Intel(R) Xeon(R) W-2133 CPU @ 3.60GHz, with an NVidia \textit{GeForce RTX} 2060 GPU. The optimisation run at $Re_D=1$ which consists of 200 iterations is chosen for evaluating the run times using different solvers, i.e. \textit{OpenFOAM}, DNN models of three sizes trained with ``Dataset-1''. 
Due to the strongly differing implementations, we compare the different solvers in terms of elapsed wall clock time.
As listed in table \ref{tab:performance}, it takes 16.3 hours using 9 cores (or 147 core-hours) for \textit{OpenFOAM} to complete such a case. Compared to \textit{OpenFOAM}, the DNN model using the GPU reduces the computational cost significantly. 
The small-scale model requires 97 seconds and even the large-scale model only takes less than 200 seconds to accomplish the task. Therefore, relative to \textit{OpenFOAM}, the speed-up factor is between 600X and 300X.
Even when considering a factor of ca. 10 in terms of GPU advantage due to an improved on-chip memory bandwidth, these measurements indicate the significant reductions in terms of runtime that can potentially be achieved by employing trained neural networks.

The time to train the DNN models varies with neural network size and the amount of training data. 
Taking ``Dataset-1'' as an example, the time of training start with 23 minutes for the small-scale model, up to 124 minutes for the large-scale model. 
When using ``Dataset-Range'' (8640 samples), it takes 4 hours and 12 minutes to train a large-scale ranged model.

Note that we aim for providing a possible choice rather than down-playing the alternatives, e.g. optimisation using discrete adjoints \citep{zhou_efficient_2016}. 
Given the potentially large one-time cost for training a model, learned approaches bear particular promise in settings where similar optimisation problems need to be solved repeatedly.
Considering the cost of data generation and training, we also believe that hybrid methods that combine deep-learning and traditional numerical methods pose very promising avenues for future algorithms. E.g., to employ learned models as fast initialisers for non-linear optimisations \citep{holl2019pdecontrol}.

\section{Concluding remarks}\label{sec:concluding_remarks}
\noindent In this paper, deep neural networks are trained to infer flowfields, and used as surrogate models to carry out shape optimisation for drag minimisation of the flow past a profile with a given area subjected to the two-dimensional incompressible fluid at low Reynolds numbers. Both level-set and B{\'e}zier curve representations are adopted to parameterise the shape, the integral values on the re-sampled Cartesian mesh are used as the optimisation objective. The gradient flow that drives the evolution of shape profile is calculated by automatic differentiation in a deep learning framework, which seamlessly integrates with trained neural network models.

Through optimisation, the drag values predicted by neural network models agree well with the \textit{OpenFOAM} results showing consistent trends. Although the total drag decreases, it is observed that the inviscid drag decreases while the contribution of viscous part increases, which is associated with the elongation of the shape. 
It is demonstrated that the present DNN model is able to predict satisfactory drag forces and the proposed optimisation framework is promising to be used for general aerodynamic design. 
Moreover, the DNN model stands out with respect to its flexibility, as it predicts a full flowfield in a region of interest and, once trained, can potentially be used in other optimization tasks with multiple objectives. In conjunction with the low run-time of the trained deep neural network, we believe the proposed method showcases the possibilities of using deep neural networks as surrogates for optimisation problems in physical sciences.

\section*{Acknowledgements} 
The authors are grateful to Oguzhan Karakaya and Hao Ma for the valuable discussions. 
This work was supported by European Research Council (ERC) grants 637014 (realFlow), and 838342 (dataFlow).

\section*{Declaration of interests} 
The authors report no conflict of interest.


\appendix
\renewcommand\thefigure{\thesection\arabic{figure}}

\renewcommand{\thetable}{\thesection\arabic{table}}

\section{Appendix}\label{appA}
\setcounter{figure}{0}    
\setcounter{table}{0}

To assess the effects of the number of samples in the datasets on the training and validation losses, six training runs are conducted with various amounts of data, which are listed in table \ref{tab-appA:dataset_details_range}. 
The method to generate those datasets has been discussed in \S\ref{subsec:dataset_generation}. Note that in table \ref{tab-appA:dataset_details_range} any smaller dataset is a subset of a larger dataset and all follow the same probability distribution (see figure \ref{fig:randomMapMethod-III}).

We found validation sets of several hundred samples to yield stable estimates, and hence use an upper limit of 400 as the maximal size of the validation dataset. The typical number of epochs, $Epoch_{max}$, for training ranges from 500 to 750.

\begin{figure}
\centering

\subfloat[Training loss]{\includegraphics[width=.5\linewidth]{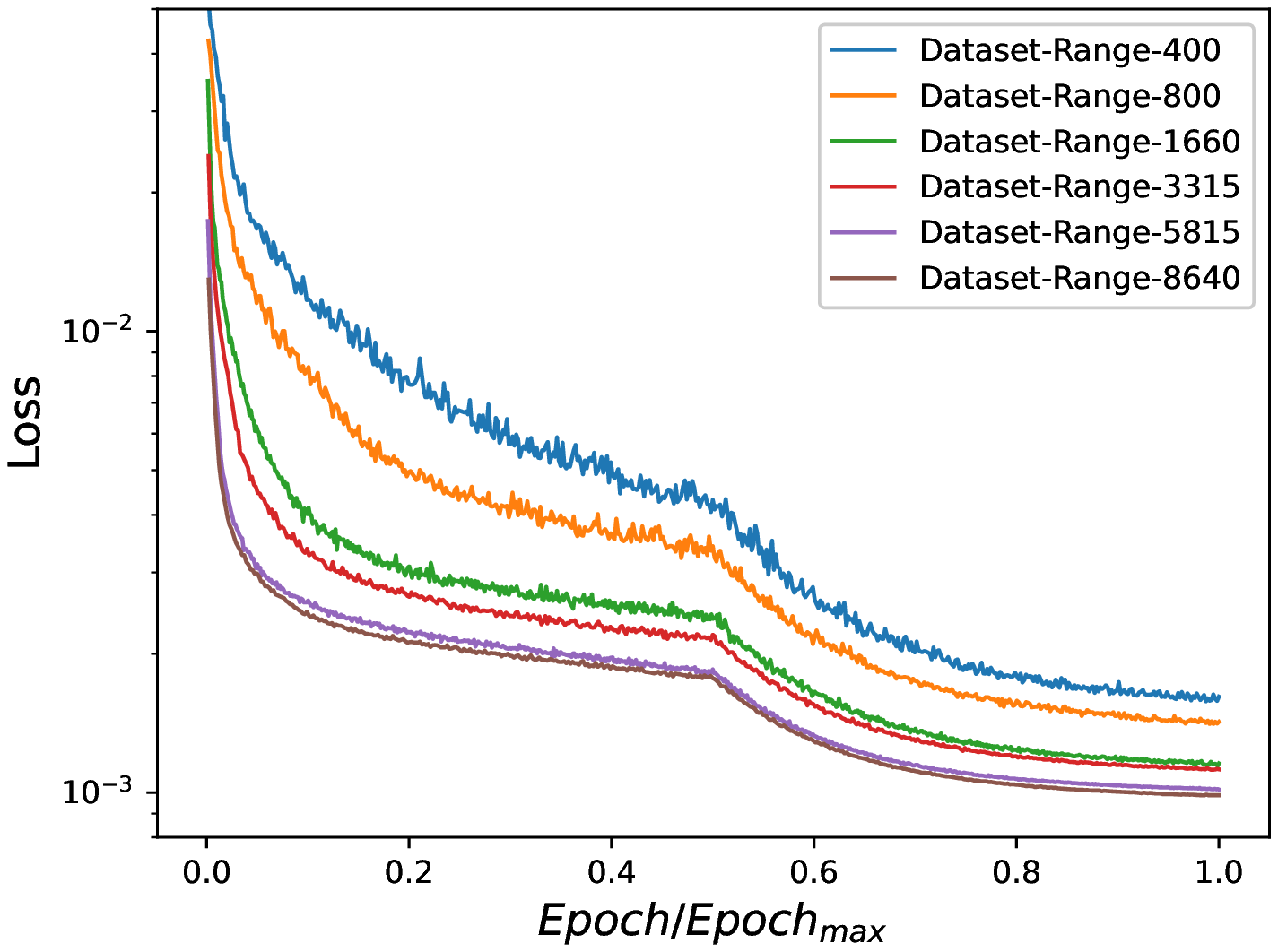}}
\subfloat[Validation loss]{\includegraphics[width=.5\linewidth]{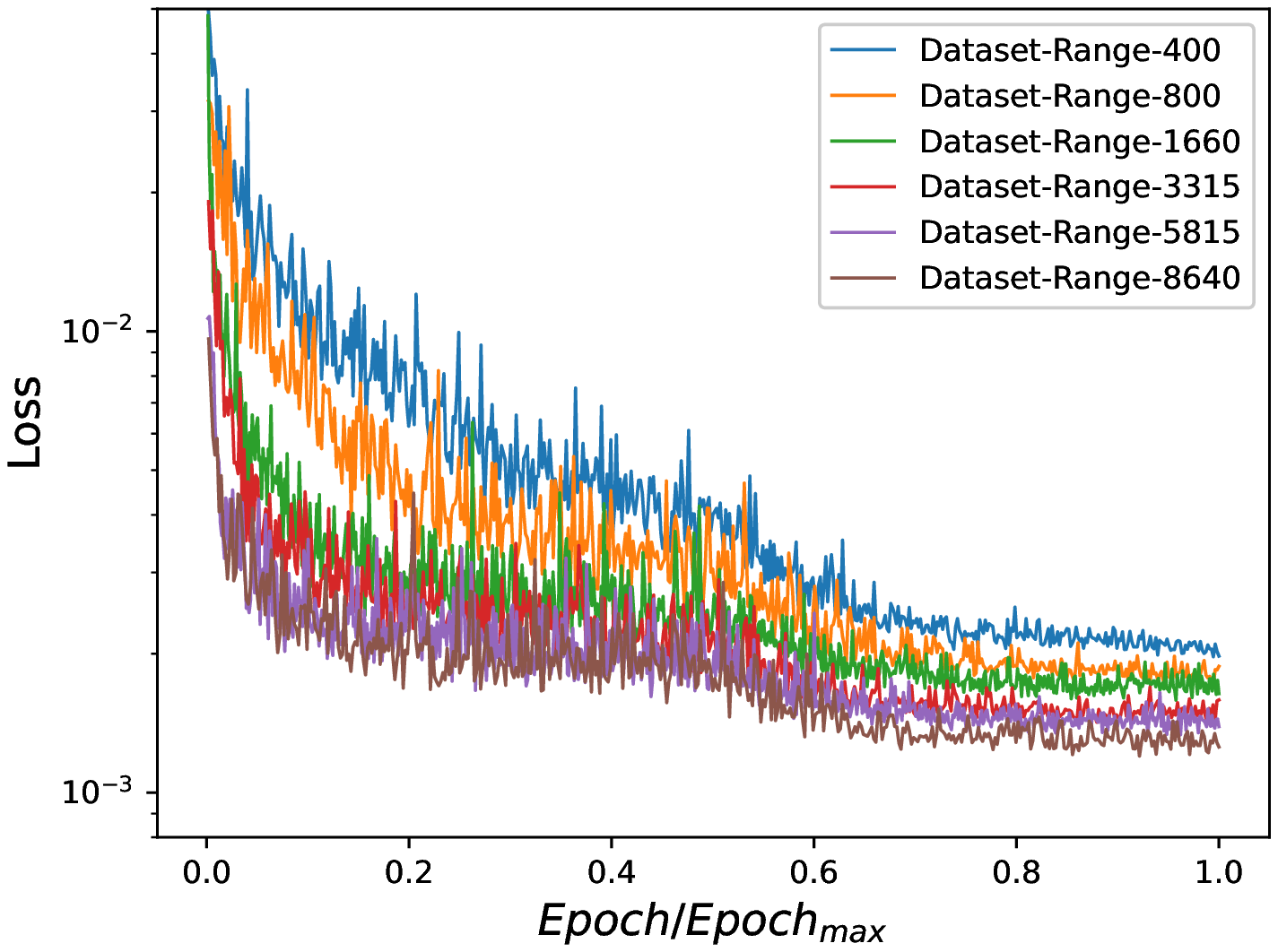}}
\caption{Comparison of training histories for different training data amounts.}
\label{fig-appA:training_losses}
\end{figure}

In figures \ref{fig-appA:training_losses}(a) and \ref{fig-appA:training_losses}(b), all models converge to stable levels of training and validation losses, and do not exhibit overfitting despite the reduced amount of data for some of the runs. For all graphs, the onset of learning rate decay can be seen in the middle of the plot. 
It can be seen that with an increased number of samples, the training and validation losses follow an overall declining trend, 
and the variance of training losses noticeably decreases. 
Additionally, looking at the models trained with ``Dataset-Range-5815'' and ``Dataset-Range-8640'', the two curves are very close in terms of the loss values, which implies that further increasing the amount of data does not yield significant improvements in terms of inference accuracy.

\begin{figure}
\centering
\includegraphics[width=0.6\textwidth]
{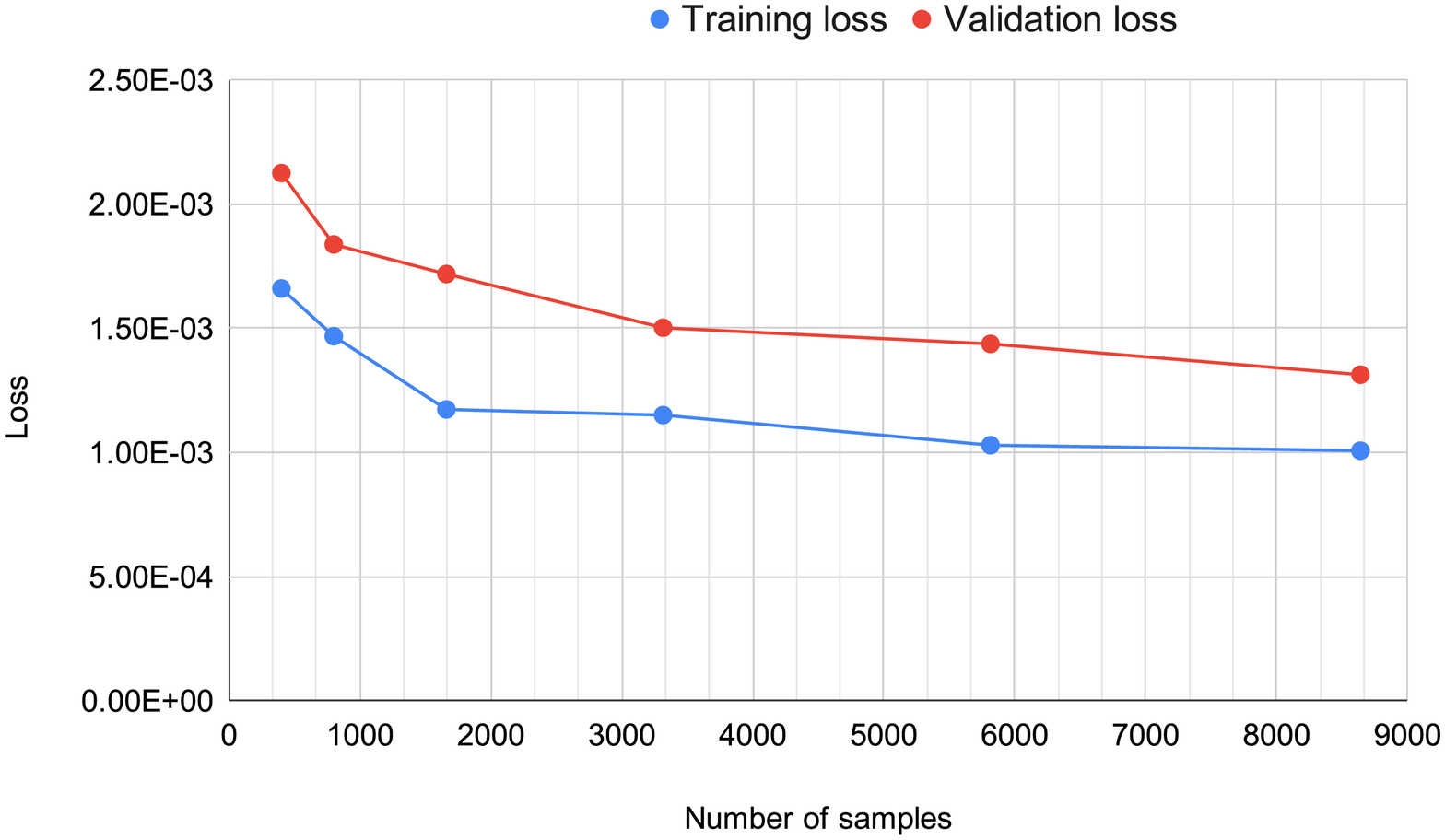}
\caption{Training and validation losses (L1) for different training data amounts.}
\label{fig-appA:numberOfSample}
\end{figure}

Figure \ref{fig-appA:numberOfSample} shows the values of training and validation losses (averaged in the last 100 epochs). 
It can be observed that the models with small amounts of data exhibit larger losses. Both training and validation loss curves 
show drastic drops over the course of the first data points with small amounts of data. 
This indicates that adding samples leads to remarkable improvements when the data amount is smaller than 3000.
The behavior stabilizes with larger amounts of data being available for training, especially when the number of samples is greater than 5000.

Based on the above test results, 
``Dataset-Range-8640'' is chosen for the optimisation study at the Reynolds number range $Re\in[0,40]$ above. 
For the sake of brevity, ``Data-Range'' is used to denote this dataset in the main text.

\begin{table}
  \begin{center}
\def~{\hphantom{0}}
  \begin{tabular}{lccc}
      \toprule
      Name    & \# of flowfields & training  & validation \\
      \midrule 
      Dataset-Range-400    & 400  & 320 & 80\\
      Dataset-Range-800    & 800  & 640 & 160\\
      Dataset-Range-1660    & 1660  & 1330 & 330\\
      Dataset-Range-3315    & 3315  & 2915 & 400\\
      Dataset-Range-5815    & 5815 & 5415 & 400\\
      Dataset-Range-8640    & 8640 & 8240 & 400\\
      \bottomrule
  \end{tabular}
  \caption{Different dataset sizes used for training runs with corresponding splits into training and validation sets.}
  \label{tab-appA:dataset_details_range}
  \end{center}
\end{table}

\bibliographystyle{unsrtnat}
\bibliography{main}
\end{document}